\documentclass{sig-alternate-05-2015}

\usepackage{times}
\usepackage{amssymb}

\usepackage{amsthm}
\usepackage{amsmath}
\usepackage{epsfig}
\usepackage{enumitem}
\makeatletter
\newif\if@restonecol
\makeatother

\usepackage[ruled, vlined]{algorithm2e}
\usepackage[para]{footmisc}
\usepackage{array}
\usepackage{comment}
\usepackage{multicol}
\usepackage{balance}
\usepackage{hyperref}
\usepackage{breakurl}
\usepackage{pgfplots}
\usepackage{multirow}
\usepackage{pbox}
\usepackage{colortbl}

\let\emptyset\varnothing

\newcommand{\eat}[1]{}

{
\theoremstyle{definition}
\newtheorem{definition}{Definition}
\newtheorem{proposition}{Proposition}
\newtheorem{theorem}{Theorem}

}

\newcommand{\entity}[1]{\textsf{\scriptsize #1}}

\newcommand{\etriple}[3]{$\langle \entity{#1},\entity{#2},\entity{#3}\rangle$}
\newcommand{\etype}[1]{\textsc{\scriptsize #1}}
\newcommand{\etypesmall}[1]{\textsc{\tiny #1}}
\newcommand{\edge}[1]{\textsf{\emph{\scriptsize #1}}}

\newcommand{\NP}{{\bf NP}\xspace}

\def\argmax{\mathop{\rm arg\ max}}

\begin{document}

\CopyrightYear{2016}
\setcopyright{acmcopyright}
\conferenceinfo{SIGMOD'16,}{June 26-July 01, 2016, San Francisco, CA, USA}
\isbn{978-1-4503-3531-7/16/06}\acmPrice{\$15.00}
\doi{http://dx.doi.org/10.1145/2882903.2915221}

\title{Generating Preview Tables for Entity Graphs}

\numberofauthors{1}
\author{
\alignauthor \textsuperscript{1}Ning Yan\titlenote{\small Work done while at the University of Texas at Arlington.}\hspace{4mm} \textsuperscript{2}Sona Hasani\hspace{4mm} \textsuperscript{2}Abolfazl Asudeh\hspace{4mm}  \textsuperscript{2}Chengkai Li\\ \vspace{1mm}
\affaddr{\textsuperscript{1}Huawei U.S. R\&D Center\hspace{4mm} \textsuperscript{2}The University of Texas at Arlington}\\ \vspace{1mm}
\email{ning.yan.uta@gmail.com\hspace{4mm} \{sona.hasani,ab.asudeh\}@mavs.uta.edu\hspace{4mm} cli@uta.edu}
}

\maketitle

\begin{abstract}
Users are tapping into massive, heterogeneous entity graphs for many applications. It is challenging to select entity graphs for a particular need, given abundant datasets from many sources and the oftentimes scarce information for them.  We propose methods to produce preview tables for compact presentation of important entity types and relationships in entity graphs. The preview tables assist users in attaining a quick and rough preview of the data.  They can be shown in a limited display space for a user to browse and explore, before she decides to spend time and resources to fetch and investigate the complete dataset.
We formulate several optimization problems that look for previews with the highest scores according to intuitive goodness measures, under various constraints on preview size and distance between preview tables.  The optimization problem under distance constraint is \NP-hard.  We design a dynamic-programming algorithm and an Apriori-style algorithm for finding optimal previews.  Results from experiments, comparison with related work and user studies demonstrated the scoring measures' accuracy and the discovery algorithms' efficiency.
\end{abstract}

\vspace{-1mm}
\section{Introduction}
\label{sec:introduction}
\begin{figure}[t]
\begin{center}
     \includegraphics[width=0.41\textwidth]{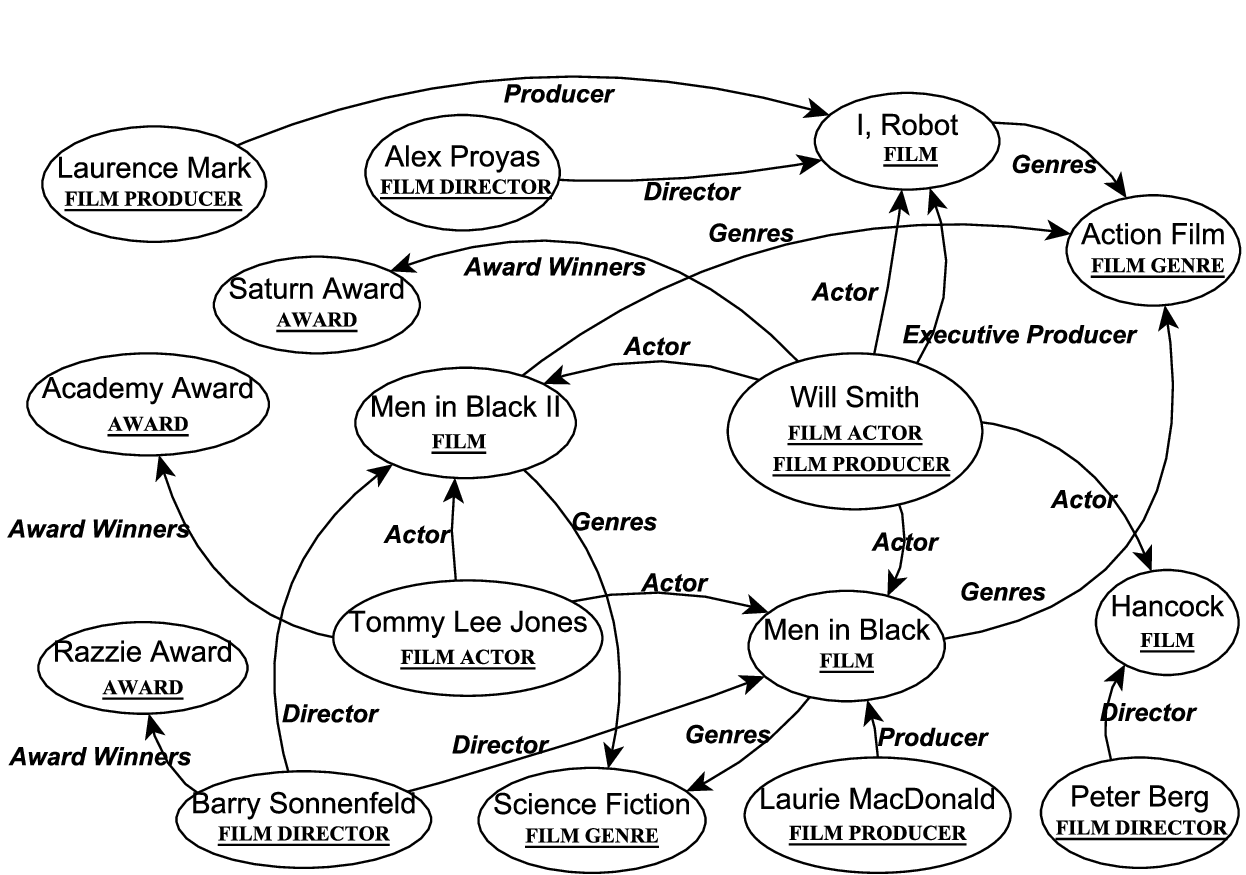}
\end{center} \vspace{-4mm}
\caption{\small An excerpt of an entity graph.} \label{fig:ergraph}
\end{figure}

We witness an unprecedented proliferation of massive, heterogeneous \emph{entity graphs} that represent entities and their relationships in many domains. For instance, in Fig.~\ref{fig:ergraph}---a tiny excerpt of an entity graph, the edge labeled \edge{Actor} between nodes \entity{Will Smith} and \entity{Men in Black} captures the fact that the person is an actor in the film.  Real-world entity graphs include knowledge bases (e.g., DBpedia~\cite{AuerBK+07}, YAGO~\cite{SuchanekKW07}, Probase~\cite{probase}, Freebase~\cite{Bollacker+08freebase} and Google's Knowledge Vault~\cite{knowledgevault}), social graphs, biomedical databases, and program analysis graphs, to name just a few. Numerous applications are tapping into entity graphs in domains such as search, recommendation systems, business intelligence and health informatics.

Entity graphs are often represented as RDF triples, due to heterogeneity of entities and the often lacking schema.  The Linking Open Data community has interlinked billions of RDF triples spanning over several hundred datasets ({\url{http://linkeddata.org}}).  Many other entity graph datasets are also available from data repositories such as the NCBI databases ({\url{http://www.ncbi.nlm.nih.gov}}), Amazon's Public Data Sets ({\url{http://aws.amazon.com/publicdatasets}}) and Data.gov ({\url{http://www.data.gov}}).

It is challenging to select entity graphs for a particular need, given abundant datasets from many sources and oftentimes scarce information available about them.  While sources such as the aforementioned data repositories often provide dataset descriptions, one cannot get a direct look at an entity graph before fetching it. Downloading a dataset and loading it into a database can be a daunting task. A data worker may need to tackle many challenges before they can start any real work on an entity graph.

In this paper, we propose methods to automatically produce \textbf{\emph{preview tables}} for entity graphs.  Given an entity graph with a large number of entities and relationships, our methods select from the many entity types a few important ones and produce a table for each chosen entity type. Such a table comprises a set of attributes, selected among many candidates, each of which corresponds to a relationship associated with the corresponding entity type.  A tuple in the table consists of an entity belonging to the entity type and its related entities for the table attributes.

Fig.~\ref{fig:preview} is a possible preview of the entity graph in Fig.~\ref{fig:ergraph}.  It consists of two preview tables---the upper table has attributes \etype{Film}, \edge{Director} and \edge{Genres}, and the lower table has attributes \etype{Film Actor} and \edge{Award Winners}.  In this preview, entities of types \etype{Film} and \etype{Film Actor} are deemed of central importance in the entity graph.  Hence, \etype{Film} and \etype{Film Actor} are the \emph{key attributes} of the two tables, respectively, marked by the underlines beneath them.  Attributes \edge{Director} and \edge{Genres} in the upper table are considered highly related to \etype{Film} entities. Similarly, \edge{Award Winners} in the lower table is highly related to \etype{Film Actor} entities.  The two tables contain 4 and 2 tuples, respectively.  For instance, the first tuple of the upper table is $t_1=$ \etriple{Men in Black}{Barry Sonnenfeld}{\{Action Film, Science Fiction\}}.  The tuple indicates that entity \entity{Men in Black} belongs to type \etype{Film} and has a relationship \edge{Director} from \entity{Barry Sonnenfeld} and has relationship \edge{Genres} to both \entity{Action Film} and \entity{Science Fiction}.

\begin{figure}\scriptsize
\begin{center}
\begin{tabular}{|c|c|c|c|}
\hline
& \textbf{\underline{\etype{Film}}} & \textbf{\edge{Director}} & \textbf{\edge{Genres}} \\
\hline
$t_1$ & \entity{Men in Black} & \entity{Barry Sonnenfeld} & \{\entity{Action Film}, \entity{Science Fiction}\} \\
\hline
$t_2$ & \entity{Men in Black II} & \entity{Barry Sonnenfeld} & \{\entity{Action Film}, \entity{Science Fiction}\} \\
\hline
$t_3$ & \entity{Hancock} & \entity{Peter Berg} & - \\
\hline
$t_4$ & \entity{I, Robot} & \entity{Alex Proyas} & \{\entity{Action Film}\} \\
\hline
\end{tabular}\vspace{1mm}

\begin{tabular}{|c|c|c|c|}
\hline
& \textbf{\underline{\etype{Film Actor}}} & \textbf{\edge{Award Winners}}  \\
\hline
$t_5$ & \entity{Will Smith} & \entity{Saturn Award} \\
\hline
$t_6$ & \entity{Tommy Lee Jones} & \entity{Academy Award} \\
\hline
\end{tabular}
\vspace{-2mm}
\caption{\small A 2-table preview of the entity graph in Fig.~\ref{fig:ergraph}. (Upper and lower tables for subgraphs \#1 and \#2 in Fig.~\ref{fig:scgraph}, respectively.)}
\label{fig:preview}
\vspace{-6mm}
\end{center}
\end{figure}

Data workers browse and explore data under inevitable display space constraints on mobile devices and desktop monitors.
The proposed preview tables are for compact presentation of important types of entities and their relationships in an entity graph.
They assist data workers in attaining a quick and rough preview of the schema of the data, before they decide to spend more time, money and
resources to fetch and investigate the complete entity graph.
The tuples in the tables further facilitate an intuitive understanding of the data.
(Our approach shows a few randomly sampled tuples in each preview table.  How to selectively display important tuples is left to future study.)

\begin{figure}[t]
\begin{center}
     \includegraphics[width=0.33\textwidth]{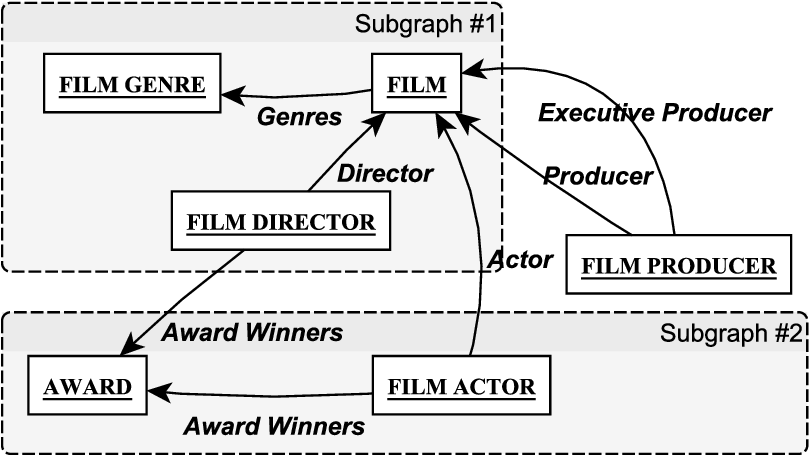}
\end{center} \vspace{-6mm}
\caption{\small The schema graph for the entity graph in Fig.~\ref{fig:ergraph}.} \label{fig:scgraph}
\end{figure}

To this end, two other approaches are arguably less adequate for gaining a quick overview of an entity graph.

(1) One solution is to show a schema graph corresponding to the data graph.  Fig.~\ref{fig:scgraph} is the schema graph for the entity graph in Fig.~\ref{fig:ergraph}.  While its definition is given in Sec.~\ref{sec:definitions}, we note that it is generated by merging same-type entity graph vertices (i.e., entities) and edges (i.e., relationships).  Although a schema graph is much smaller than the corresponding entity graph, it is not small enough for easy presentation and quick preview.  For instance, in a snapshot of the ``film'' domain of Freebase, there are 190K vertices and 1.6M edges.  The corresponding schema graph consists of 50 entity types and 136 relationship types.

(2) Another approach is to present a summary of the schema graph, by schema summarization techniques~\cite{DBLP:journals/pvldb/YangPS09,DBLP:journals/pvldb/YangPS11a,DBLP:conf/vldb/YuJ06a,DBLP:conf/sigmod/TianHP08,DBLP:conf/icde/ZhangTP10}. Some of these methods~\cite{DBLP:journals/pvldb/YangPS09,DBLP:journals/pvldb/YangPS11a,DBLP:conf/vldb/YuJ06a} work on relational and semi-structured data, instead of graph data.
Some~\cite{DBLP:conf/vldb/YuJ06a,DBLP:conf/sigmod/TianHP08,DBLP:conf/icde/ZhangTP10} produce trees or graphs as output instead of flat tables.
It is unclear how to apply these methods on an entity graph or its schema graph, due to differences in data models.
Although it is plausible that some of these approaches can be adapted for entity graphs, several reasons can render them ineffective. \emph{First}, schema summary can still be quite large.  The most closely related work, \cite{DBLP:journals/pvldb/YangPS09,DBLP:journals/pvldb/YangPS11a}, clusters the tables in a database but does not reduce the number of tables or the complexity of database schema.  If we treat each entity type as a table and its neighboring entity types in the schema graph as the table attributes, the number of tables would equal the number of entity types.  For the aforementioned ``film'' domain in Freebase, it means one would have to understand the result of clustering 50 tables.  \emph{Second}, schema summarization is for helping database administrators and programmers in gaining a detailed understanding of a database in order to form queries.  Our goal is to assist data workers in attaining a quick and rough understanding of an entity graph, before they decide to grasp such a detailed understanding.  Therefore, our work can be viewed as an approach of finding a succinct representation of the schema graph (instead of clustering it). We are not aware of such an approach in previous studies.

In our definition (details in Sec.~\ref{sec:definitions}), a \emph{preview} is a set of preview tables, each of which has a \emph{key attribute} (corresponding to an entity type) and a set of \emph{non-key attributes} (each corresponding to a relationship type).  Given an entity graph and its schema graph, there is thus a large space of possible previews.  Our goal is to find an ``optimal'' preview in the space.  To this end, we tackle several challenges:  (1) We discern what factors contribute to the goodness of a preview and propose several scoring functions for key and non-key attributes as well as preview tables. The scoring functions are based on several intuitions related to how much information a preview conveys and how helpful it is to users. (2) Based on the scoring measures, a preview's score is maximized when it includes as many tables and attributes as possible.  However, the purpose of having a preview is to help users attain a quick understanding of data and thus a preview must fit into a limited display space.  Considering the tradeoff, we enforce a constraint on preview size.  Furthermore, we consider enforcing an additional constraint on the pairwise distance between preview tables.  Given the spaces of all possible previews, we formulate the optimization problem of finding an preview with the highest score among those satisfying the constraints.  The optimization is non-trivial, as we prove that it is \NP-hard under distance constraint.  (3) The search space of previews grows exponentially by data size and the constraints.  A brute-force approach is thus too costly.  For efficiently finding optimal previews, we designed a dynamic programming algorithm and an Apriori~\cite{apriori}-style algorithm.

In summary, this paper makes the following contributions: \vspace{-1mm}
\begin{list}{$\bullet$}
{ \setlength{\leftmargin}{1em} \setlength{\parsep}{-3.2pt}}
  \item We motivated a novel concept of preview for entity graphs.
  \item We proposed ideas for measuring the goodness of previews based on several intuitions. (Sec.~\ref{sec:scoring})
  \item We formulated optimal preview discovery problem, and proved its \NP-hardness under distance constraint. (Sec.~\ref{sec:opt})
  \item We developed a dynamic-programming algorithm and an Apriori-style algorithm for finding optimal previews. (Sec.~\ref{sec:algorithms})
  \item Extensive experiments, comparison with related work, and user study verified the scoring measures' accuracy, the algorithms' efficiency, and the effectiveness of discovered previews. (Sec.~\ref{sec:experiments})
\end{list}

\section{Preview Discovery Problem}
\label{sec:definitions}

An \emph{entity graph} is a directed multigraph $G_d(V_d,E_d)$ with vertex set $V_d$ and edge set $E_d$.
Each vertex $v \in V_d$ represents an entity and each edge $e(v, v') \in E_d$ represents a directed relationship from entity $v$ to $v'$.
$G_d$ is a multigraph since there can be multiple edges between two vertices.
(E.g., in Fig.~\ref{fig:ergraph}, there are two edges \edge{Actor} and \edge{Executive Producer} from entity \entity{Will Smith} to entity \entity{I, Robot}.)

\begin{table}[t]
\scriptsize
\centering
\begin{tabular}{|@{\hspace{0.3em}}c@{\hspace{0.3em}}|>{\hspace{-1mm}}p{54mm}|}
\hline
$G_d(V_d,E_d)$ & an entity graph \\
\hline
$v \in V_d$ &  an entity \\
\hline
$e(v, v') \in E_d$ &  a directed relationship from entity $v$ to entity $v'$ \\
\hline
$G_s(V_s,E_s)$ & a schema graph \\
\hline
$\tau \in V_s$ & an entity type \\
\hline
$\gamma(\tau, \tau') \in E_s$ & a relationship type from entity type $\tau$ to entity type $\tau'$ \\
\hline
$T$ & a preview table \\
\hline
$T.key$ & the key attribute of $T$ \\
\hline
$T.nonkey$ & the non-key attributes of $T$ \\
\hline
$T.\tau$ & the set of entities of type $\tau$---the key attribute of $T$ \\
\hline
$t \in T$ & a tuple $t$ in preview table $T$ \\
\hline
$t.\tau$ & $t$'s value on $\tau$ which is the key attribute of $T$ \\
\hline
$t.\gamma$ & $t$'s value on non-key attribute $\gamma$ \\
\hline
$\mathcal{P} = \{\mathcal{P}[1], ..., \mathcal{P}[k]\}$ & a preview, which consists of $k$ preview tables\\
\hline
$\mathcal{P}_{opt}$ & an optimal preview\\
\hline
$S(\mathcal{P})$ & the score of preview $\mathcal{P}$\\
\hline
$S(T)$ & the score of preview table $T$\\
\hline
$S_{cov}(\tau)$, $S_{walk}(\tau)$ & score of key attribute $\tau$ based on coverage/random-walk\\
\hline
$S^{\tau}_{cov}(\gamma)$, $S^{\tau}_{ent}(\gamma)$  & score of non-key attribute $\gamma$ based on coverage/entropy  \\
\hline
$\mathbb{T}$ & the space of all possible preview tables\\
\hline
$\mathbb{P}$ & the space of all possible previews\\
\hline
$dist(\tau, \tau')$ & distance between $\tau$ and $\tau'$ in schema graph $G_s$\\
\hline
\end{tabular}
\vspace{-3mm}
\caption{\small Notations.}\label{tab:notation}\vspace{-1mm}
\end{table}

Each entity is labeled by a name.  For simplicity and intuitiveness of presentation, we shall mention entities by their names, assuming all entities have distinct names, although in reality they are distinguished by unique identifiers such as URIs.  Each entity belongs to one or more \emph{entity types}, underlined in Fig.~\ref{fig:ergraph}.  (E.g., \entity{Will Smith} belongs to types \etype{Film Actor} and \etype{File Producer} and \entity{I, Robot} belongs to type \etype{Film}.)  Each relationship belongs to a \textit{relationship type}. (E.g., the edge from \entity{Will Smith} to \entity{Men in Black} has type \edge{Actor}.)  The type of a relationship determines the types of its two end entities.  For instance, an edge of type \edge{Actor} is always from an entity belonging to \etype{File Actor} to an entity belonging to \etype{Film}.  We will mention edges by the surface names of their relationship types.  Two different relationship types may have the same surface name for intuitively expressing their meanings, although underlyingly they have different identifiers.  For instance, the \edge{Award Winners} edge from \entity{Will Smith} to \entity{Saturn Award} and the \edge{Award Winners} edge from \entity{Barry Sonnenfeld} to \entity{Razzie Award} belong to two different relationship types.  The former is for relationships from \etype{Film Actor} to \etype{Award}, while the latter is for relationships from \etype{Film Director} to \etype{Award}.

Given an entity graph $G_d(V_d,E_d)$, its \emph{schema graph} is a directed graph $G_s(V_s,E_s)$,
where each vertex $\tau \in V_s$ represents an entity type and each directed edge
$\gamma(\tau, \tau') \in E_s$ represents a relationship type from entity type $\tau$ to $\tau'$.
An edge $\gamma(\tau, \tau') \in E_s$ if and only if there exists an edge $e(v, v') \in E_d$
where $e$ has type $\gamma$, $v$ has type $\tau$ and $v'$ has type $\tau'$.
Fig.~\ref{fig:scgraph} shows the schema graph corresponding to the entity graph in Fig.~\ref{fig:ergraph}.
A schema graph is a multigraph as there can be multiple relationship types between two entity types.
(E.g., two relationship types---\edge{Producer} and \edge{Executive Producer}---are from entity type \etype{Film Producer} to \etype{Film}.)
It is clear from the above definitions that, given a data graph, the corresponding schema graph is uniquely determined.\vspace{-1mm}

\begin{definition}[Preview Table and Preview] \label{def:preview}
Given an entity graph $G_d(V_d,E_d)$ and its schema graph $G_s(V_s,E_s)$,
a \emph{preview table} $T$ has a mandatory \emph{key attribute} (denoted $T.key$)
and at least one \emph{non-key attributes} (denoted $T.nonkey$).
$T$ corresponds to a star-shape subgraph of the schema graph $G_s(V_s,E_s)$.
The key attribute corresponds to an entity type $\tau \in V_s$, and each non-key attribute
corresponds to a relationship type $\gamma(\tau, \tau') \in E_s$ or $\gamma(\tau', \tau) \in E_s$.
Note that the edges from and to an entity are both important.
Hence, a non-key attribute corresponds to either $\gamma(\tau, \tau')$ or $\gamma(\tau', \tau)$.

The preview table $T$ consists of a set of tuples.
The number of tuples equals the number of entities of type $\tau$ (the key attribute of $T$),
i.e., $|T| = |T.\tau|$ and $T.\tau = \{v|v \in V_d \wedge v \text{ has type } \tau\}$.
Given an arbitrary tuple $t \in T$, we denote $t$'s key attribute value by $t.\tau$.
Each tuple $t$ attains a distinct value of $t.\tau$.
Its value on a non-key attribute $\gamma(\tau, \tau')$, denoted $t.\gamma(\tau, \tau')$ or simply $t.\gamma$,
is a set---the set of entities in entity graph $G_d$ incident from $t.\tau$ through an
edge of type $\gamma(\tau, \tau')$.
More formally, $t.\gamma(\tau, \tau') = \{u| u \in V_d \wedge e(t.\tau,u) \in E_d \wedge u \text{ belongs to type } \tau'\}$.
Symmetrically, its value on a non-key attribute $\gamma(\tau', \tau)$ is the set of entities in $G_d$ incident to $t.\tau$ through an edge of type $\gamma(\tau', \tau)$,
i.e., $t.\gamma(\tau', \tau) = \{u| u \in V_d \wedge e(u,t.\tau) \in E_d \wedge u \text{ belongs to type } \tau'\}$.

A \emph{preview} $\mathcal{P}$ is a set of preview tables, i.e., $\mathcal{P}=\{\mathcal{P}[1], ..., \mathcal{P}[k]\}$,
where $\forall i\neq j, \mathcal{P}[i].key \neq \mathcal{P}[j].key$,
$k \leqslant |V_s|$ is the total number of preview tables.
Note that $|V_s|$ is the number of vertices in $G_s$, i.e., the number of entity types in $G_d$. \qed\vspace{-2mm}
\end{definition}

According to Definition~\ref{def:preview}, the upper and lower tables in Fig.~\ref{fig:preview} correspond to the star-shape subgraphs \#1 and \#2 in Fig.~\ref{fig:scgraph}, respectively.
The key attribute in the upper table is \etype{Film} and the non-key attributes are \edge{Director} and \edge{Genres}.
The key attribute in the lower table is \etype{Film Actor} and its non-key attribute is \edge{Award Winners}.
It is worth noting that, although each tuple's value on the key attribute is non-empty, unique and single-valued, its value on a non-key attribute can be empty (e.g., $t_3.\edge{Genres}$ in Fig.~\ref{fig:preview}), duplicate
(e.g., $t_1.\edge{Director}$ and $t_2.\edge{Director}$ in Fig.~\ref{fig:preview}) and multi-valued (e.g., $t_1.\edge{Genres}$ and $t_2.\edge{Genres}$ in Fig.~\ref{fig:preview}).  It also follows that a preview table is not a relational table.

By Definition~\ref{def:preview}, every vertex $\tau$ in a schema graph can serve as the key attribute of a candidate preview table, which also includes at least one non-key attribute---an edge incident on $\tau$.  We use $\mathbb{T}$ to denote the space of all possible preview tables.  A preview is a set of preview tables.  We use $\mathbb{P}$ to denote the space of all possible previews.  Note that $\mathbb{P} \subset 2^\mathbb{T}$, i.e., not every member of the power set $2^\mathbb{T}$ is a valid preview, because by Definition~\ref{def:preview} preview tables in a preview cannot have the same key attribute.\vspace{-2mm}

{\flushleft \textbf{Problem Statement}}: Given an entity graph $G_d(V_d,E_d)$ and its corresponding schema graph $G_s(V_s,E_s)$, the \emph{preview discovery problem} is to find $\mathcal{P}_{opt}$---the optimal preview among all possible previews.
We shall develop the notions of goodness and optimality for a preview and define goodness measures in Sec.~\ref{sec:scoring}.

Note that the preview discovery problem focuses on selecting key and non-key
attributes for preview tables.  It does not select tuples.
As our goal is to help users attain a good initial understanding of the
schema of an entity graph, we argue that it is only necessary to show a small number of tuples
instead of all.  Our current approach is to randomly select a few.
How to choose the most representative tuples is left for future work.

\vspace{-1mm}

\section{Scoring Measures for Previews}
\label{sec:scoring}

\setlength{\belowdisplayskip}{0pt} \setlength{\belowdisplayshortskip}{0pt}
\setlength{\abovedisplayskip}{0pt} \setlength{\abovedisplayshortskip}{0pt}

In this section, we discuss the scoring functions for measuring the
goodness of previews for entity graphs.  The measures are based on
the intuition that a good preview should 1) relate to as many entities
and relationships as possible and 2) help users understand an entity
graph and its schema graph.  The first intuition is obvious, as a
preview relating to only a small number of entities or relationships will
inevitably lose lots of information and thus lead to poor comprehensibility
of the original graph.  The second intuition models the goodness of previews
according to users' behavior in browsing entity and schema graphs.\vspace{-1mm}

\subsection{Preview Scoring}

The score of a preview $\mathcal{P}=\{\mathcal{P}[1], ..., \mathcal{P}[k]\}$
is simply aggregated from individual preview tables' scores, by summation:\vspace{-1mm}
\begin{equation}\label{eq:preview-score}
S(\mathcal{P}) = \sum_{i=1}^{k} S(\mathcal{P}[i]),
\end{equation}
where $S(\mathcal{P}[i])$ is the score of a preview table $\mathcal{P}[i]$, defined as:\vspace{-1mm}
\begin{equation}\label{eq:table-score}
S(\mathcal{P}[i]) = S(\tau) \times \!\!\!\!\! \sum_{\gamma \in \mathcal{P}[i].nonkey} \!\!\!\!\!\!\!\!\!\! S^{\tau}(\gamma),
\end{equation}
where $S(\tau)$ is the score of the key attribute of $\mathcal{P}[i]$ (i.e., $\mathcal{P}[i].key$=$\tau$)
and $S^{\tau}(\gamma)$ is the score of a non-key attribute $\gamma$ in $\mathcal{P}[i]$.
$S(\tau)$ and $S^{\tau}(\gamma)$ are defined and elaborated in
Sec.~\ref{sec:key-score} and Sec.~\ref{sec:non-key-score}.

In the above definition, the score of a preview table equals the product of its key attribute's score and the summation of its non-key attributes' scores.  The definition gives the key attribute $\tau$ much higher importance than any individual non-key attribute, because the preview table centers around the entities of type $\tau$ and describes their non-key attributes, i.e., their relationships with other entities.

It is possible to propose many viable scoring functions for previews, key attributes and non-key attributes.
Furthermore, techniques such as learning-to-rank~\cite{learn-to-rank}
may be applied in ranking previews by features related to key
and non-key attributes, although the feasibility of collecting many labelled
data is less clear in this case.  We leave it to future work to explore this direction.
Nevertheless, we note that the results on the optimization problems in Section~\ref{sec:opt}
and the algorithms in Section~\ref{sec:algorithms} will stand, as long as the scoring
function replacing Eq.~\ref{eq:preview-score} and Eq.~\ref{eq:table-score} is
monotonic with regard to $S(\tau)$ and $S^{\tau}(\gamma)$, and the measures defining
$S(\tau)$ and $S^{\tau}(\gamma)$ do not affect the results.

\subsection{Key Attribute Scoring}\label{sec:key-score}
\vspace{-3mm}
{\flushleft \textbf{Coverage-based scoring measure:}} \hspace{2mm}
Given an entity graph $G_d(V_d$, $E_d)$ and its corresponding schema graph $G_s(V_s,E_s)$, the key attribute $\tau$ of a candidate preview table $T$ corresponds to an entity type, i.e., $\tau \in V_s$.  If the entity graph consists of many entities of type $\tau$, including $T$ in the preview makes the preview relevant to all those entities.  The coverage-based scoring measure thus defines the score of $\tau$ as the number of entities bearing that type:
\vspace{1mm}
$$
S_{cov}(\tau) = |\{v| v \in V_d \wedge v \text{ has type } \tau\}|
$$

For example, given the entity graph in Fig.~\ref{fig:ergraph} and the corresponding schema graph in Fig.~\ref{fig:scgraph}, the coverage-based score of the key attribute \etype{Film} is $S_{cov}(\etype{Film})=4$.

\vspace{-2mm}
{\flushleft \textbf{Random-walk based scoring measure:}} \hspace{2mm}
We consider a \textit{random-walk process} over a graph $G$ converted from the schema graph $G_s(V_s,E_s)$, inspired by the PageRank algorithm~\cite{Brin:1998:ALH:297805.297827} for Web page ranking and many related ideas. Similar to $G_s$, vertices in $G$ are entity types and edges are relationship types. Different from $G_s$, the edges are undirected. As explained in Def.~\ref{def:preview}, the edges from and to an entity are both important to the entity. The edge between $\tau_i$ and $\tau_j$ in $G$ is weighted by the number of relationships (i.e., the number of edges) in the entity graph between entities of types $\tau_i$ and $\tau_j$. We denote the weight by $w_{ij}$, defined as follows.
\begin{align*}
w_{ij} = w_{ji} &= \!\! \sum_{\gamma(\tau_i,\tau_j)\in E_s} \!\!\!\!\! |\{e| e \in E_d \wedge e \text{ has type } \gamma(\tau_i,\tau_j)\}| \\ \vspace{-1mm}
&+ \!\! \sum_{\gamma(\tau_j,\tau_i)\in E_s} \!\!\!\!\! |\{e| e \in E_d \wedge e \text{ has type } \gamma(\tau_j,\tau_i)\}|
\end{align*}

In the $|V_s| \times |V_s|$ \textit{transition matrix} $M$, an element $M_{ij}$ corresponds to the \textit{transition probability} from $\tau_i$ to $\tau_j$ in $G$.  $M_{ij}$ equals the ratio of $w_{ij}$ to the total weight of all edges incident on $\tau_i$ in $G$:
$$
M_{ij} = w_{ij} / \sum_{k} w_{ik}
$$

For example, based on Fig.~\ref{fig:scgraph}, the transition probability from \etype{Film} to \etype{Film Genre} is $M_{\etypesmall{Film}, \etypesmall{Film Genre}}$ = $w_{\etypesmall{Film},\etypesmall{Film Genre}} / (w_{\etypesmall{Film},\etypesmall{Film Genre}}$ + $w_{\etypesmall{Film},\etypesmall{Film Actor}}$ + $w_{\etypesmall{Film},\etypesmall{Film Director}}$ + $w_{\etypesmall{Film},\etypesmall{Film Producer}})$ = $5/(5+6+4+3)$ = $0.28$.  The transition probability from \etype{Film} to \etype{Film Producer} is $M_{\etypesmall{Film}, \etypesmall{Film Producer}}$ = $w_{\etypesmall{Film},\etypesmall{Film Producer}} / (w_{\etypesmall{Film},\etypesmall{Film Genre}}$ + $w_{\etypesmall{Film},\etypesmall{Film Actor}}$ + $w_{\etypesmall{Film},\etypesmall{Film Director}}$ + $w_{\etypesmall{Film},\etypesmall{Film Producer}})$ = $3/(5+6+4+3)$ = $0.17$.

Suppose a random walker traverses in $G$, either by going from an entity type $\tau_i$ to another entity type $\tau_j$ through the edge between them with probability $M_{ij}$ or by jumping to a random entity type.  Entity types that are more likely to be visited by the user are of higher importance.  The random walk process will converge to a stationary distribution which represents the chances of entity types being visited.  The stationary distribution $\pi$ of the random walk process is given as follows.  Note that a similar idea was applied in~\cite{DBLP:journals/pvldb/YangPS09} for ranking relational tables by importance.
$$
\pi = \pi M
$$

The random-walk based score of a candidate key attribute $\tau_i$ is:
$$
S_{walk}(\tau_i) = \pi_i, \text{where } \pi_i \text{ is the stationary probability of } \tau_i.
$$

\subsection{Non-Key Attribute Scoring}\label{sec:non-key-score}
\vspace{-3mm}
{\flushleft \textbf{Coverage-based scoring measure:}} \hspace{2mm}
The coverage-based scoring measure for non-key attribute is similar to that for key attribute.  Given an entity graph $G_d(V_d,E_d)$ and its schema graph $G_s(V_s,E_s)$, consider a candidate preview table $T$ with key attribute $\tau$.  A non-key attribute $\gamma$ of $T$ corresponds to a relationship type, i.e., $\gamma \in E_s$.  If the entity graph contains many edges (i.e., relationships) belonging to type $\gamma$, incorporating such a relationship type into table $T$ makes it relevant to all those relationships and their corresponding entities.  The coverage-based scoring measure thus defines the score of $\gamma$ as the number of relationships bearing that type:\vspace{1mm}
$$
S^{\tau}_{cov}(\gamma) = |\{e| e \in E_d \wedge e \text{ has type } \gamma\}|
$$

For example, given the entity graph in Fig.~\ref{fig:ergraph} and the schema graph in Fig.~\ref{fig:scgraph}, the coverage-based scores of non-key attributes \edge{Director} and \edge{Genres} are $S^{\etype{Film}}_{cov}(\edge{Director})=4$ and $S^{\etype{Film}}_{cov}(\edge{Genres})=5$.

The coverage-based scoring measure for non-key attribute is symmetric, i.e., given $\gamma(\tau, \tau')$ (or $\gamma(\tau', \tau)$) $\in$ $T.nonkey$, $S^{\tau}_{cov}(\gamma) \equiv S^{\tau'}_{cov}(\gamma)$.  Both $\tau$ and $\tau'$ can be the key attribute of a different preview table, in which $\gamma$ is a non-key attribute.  The scores of $\gamma$ in the two tables are equal.

\vspace{-2mm}
{\flushleft \textbf{Entropy-based scoring measure:}} \hspace{2mm}
For a preview table $T$ with key attribute $\tau$, we measure the goodness of a non-key attribute $\gamma(\tau, \tau')$ (or $\gamma(\tau', \tau)$) by how much information it provides to $T$, for which the \emph{entropy} of $\gamma$ ($H(\gamma)$) is a natural choice of measure:
$$
S^{\tau}_{ent}(\gamma) = H(\gamma) = \sum_{j=1} \frac{n_j}{|t.\gamma|} \log(\frac{|t.\gamma|}{n_j}),
$$
where $n_j$ is the number of tuples in $T$ that attain the same $j$th attribute value $u$ on
non-key attribute $\gamma(\tau, \tau')$ (or $\gamma(\tau', \tau)$), i.e.,
$u \in V_d \wedge u \text{ has type } \tau'$ and $n_j = |\{ v | v \in T.\tau \wedge e(v,u) \in E_d \text{ (or } e(u,v) \in E_d \text{)} \wedge e \text{ has type } \gamma \}|$.
$|t.\gamma|$ is the number of tuples in $T$ with non-empty values on $\gamma(\tau, \tau')$ (or $\gamma(\tau', \tau)$).
Continue the running example. The entropy-based scores of non-key attributes \edge{Director} and \edge{Genres} are $S^{\etype{Film}}_{ent}(\edge{Director})=(2/4)\log(4/2)+(1/4)\log(4/1)+(1/4)\log(4/1)=0.45$, and $S^{\etype{Film}}_{ent}(\edge{Genres})=(2/3)\log(3/2)+(1/3)\log(3/1)=0.28$.
Note that for two values on a multi-valued attribute (e.g., \{\entity{Action Film}, \entity{Science Fiction}\} and \{\entity{Action Film}\} for \etype{Film}.\edge{Genres} in Fig.~\ref{fig:preview}), we consider them equivalent if and only if they have the same set of component values.
By definition, the entropy-based scoring measure for non-key attribute is asymmetric, i.e., given $\gamma(\tau, \tau')$ (or $\gamma(\tau', \tau)$) $\in$ $T.nonkey$, $S^{\tau}_{ent}(\gamma) \not\equiv S^{\tau'}_{ent}(\gamma)$.

\section{Optimal Previews under Size and Distance Constraints}
\label{sec:opt}

In this section, based on the scoring measures defined in Sec.~\ref{sec:scoring}, we formulate several optimization problems that look for the optimal previews with best scores under various constraints on preview size and distance between preview tables.  We prove that some of these optimization problems are \NP-hard.

By Eq.~\ref{eq:preview-score} (or any other monotonic aggregate function), the score of a preview monotonically increases by its member preview tables---the more preview tables in a preview, the higher its score.  Similarly by Eq.~\ref{eq:table-score}, the score of a preview table monotonically increases by its non-key attributes.  The properties are formally stated in the following two propositions.  Recall that $\mathbb{P}$ and $\mathbb{T}$ denote the space of all possible previews and all possible preview tables.\vspace{-2mm}

\begin{proposition}\label{prop:monotone-preview}
Given previews $\mathcal{P}_1, \mathcal{P}_2 \in \mathbb{P}$, if $\mathcal{P}_1 \supseteq \mathcal{P}_2$, then $S(\mathcal{P}_1)\geq S(\mathcal{P}_2)$.\vspace{-2mm}
\end{proposition}

\begin{proposition}\label{prop:monotone-table}
Given preview tables $T_1, T_2 \in \mathbb{T}$, if $T_1.key = T_2.key$ and $T_1.nonkey \supseteq T_2.nonkey$, then $S(T_1)\geq S(T_2)$.\vspace{-2mm}
\end{proposition}

By the above propositions, a preview's score is maximized when it includes as many tables and attributes as possible.
However, a preview must fit into a limited display space, due to constraints posed by mobile devices and desktop monitors.
Therefore the size and the goodness score of a preview present a tradeoff.
Considering the tradeoff, we enforce a constraint on preview size, given by a pair of integers $(k,n)$,
where $k$ is the number of allowed preview tables and $n$ is the number of allowed
non-key attributes in the tables.  Their values may be either manually chosen by interactive users
or automatically suggested based on the size of a display space.
The previews satisfying the size constraint are called \emph{concise previews}.

An alternative size constraint is a maximally allowed number of attributes per preview table.
However, we do not consider such a constraint in this paper.  We argue that forcing each preview table to have
the same width can cause two problems---on the one hand, the allocated space for some preview tables may be wasted
because they do not have as many important non-key attributes; on the other hand, the fixed space is insufficient
for other preview tables with more important non-key attributes.

Further, for obtaining either a coherent or a diverse preview,
we enforce an additional constraint on the
pairwise distance between preview tables.  The distance between two
preview tables $T_1$ and $T_2$ (denoted $dist(T_1, T_2)$) is the length
of the shortest undirected path\footnote{\small An undirected path in a
directed graph is a path in which the edges are not all oriented in the
same direction.} between their key attributes $T_1.key$ and $T_2.key$ in
schema graph $G_s$.  (Recall that the key attributes are vertices
(i.e., entity types) in $G_s$.)  For example, the distance between the
two tables in Fig.~\ref{fig:preview} is $1$, which is the shortest path
length between \etype{Film} and \etype{Film Actor} in the schema graph in
Fig.~\ref{fig:scgraph}.  Similarly, for the two tables whose key attributes
are \etype{Film} and \etype{Award}, their distance would be $2$.

Based on the above notion of distance, the constraint on table distance is given by an integer $d$, which is the maximum (resp. minimum) distance between preview tables.  The previews satisfying the distance constraint are called \emph{tight (resp. diverse) previews}.  Intuitively speaking, the preview tables in a tight preview are highly related to each other due to their short pairwise distance, while the preview tables in a diverse preview are not tightly related to each other and cover different types of concepts.  Arguably, both types of previews are useful for understanding an entity graph.  We shall compare them empirically in Sec.~\ref{sec:experiments}.

Below we formally define the three types
of previews and the corresponding optimization problems.
Note that we assume the constraints $k,n,d$ are given.
While it is intuitive for a user to specify desired values for these
constraints, it is helpful if a system can automatically suggest
values.  We leave it to future work.

\begin{definition}[Concise, Tight and Diverse Previews]
Given the size constraint $(k,n)$, a \emph{concise preview} has $k$ preview tables (i.e., key attributes) and no more than $n$ non-key attributes in the tables.~\footnote{\small A preview with less than $n$ non-key attributes may outscore another preview with exactly $n$ non-key attributes. Further, a set of $k$ entity types may have only less than $n$ edges in the schema graph. Hence, the condition $|\mathcal{P}[i].nonkey| \leq n$ instead of $|\mathcal{P}[i].nonkey| = n$.  On the other hand, it is safe to assume that an entity graph with practical significance always has more than $k$ entity types under any reasonably small $k$.  Therefore an optimal preview always should have exactly $k$ preview tables, given the monotonic scoring function (cf. Eq.~\ref{eq:preview-score}).} The space of all concise previews is
\begin{equation}
\mathbb{P}_{k,n} = \{ \mathcal{P}\ \big|\ \mathcal{P}\in \mathbb{P}, |\mathcal{P}| = k, \sum_{i=1}^{k} |\mathcal{P}[i].nonkey| \leq n\}.\nonumber
\end{equation}

Given the size constraint $(k,n)$ and the distance constraint $d$, a \emph{tight preview} (\emph{diverse preview}) is a concise preview in which the distance between any pair of preview tables is smaller (greater) than or equal to $d$.   The space of all tight previews is
\begin{equation}
\mathbb{P}_{k,n,\leq d} = \{ \mathcal{P}\ \big|\ \mathcal{P}\in \mathbb{P}_{k,n}, \forall T_1, T_2 \in \mathcal{P}, dist(T_1,T_2) \leq d\}.\nonumber
\end{equation}
The space of all diverse previews is
\begin{equation}
\mathbb{P}_{k,n,\geq d} = \{ \mathcal{P}\ \big|\ \mathcal{P}\in \mathbb{P}_{k,n}, \forall T_1, T_2 \in \mathcal{P}, dist(T_1,T_2) \geq d\}.\nonumber \qed \vspace{-1mm}
\end{equation}
\end{definition}

Given the spaces of concise, tight and diverse previews, we
formulate three optimization problems---finding an
\emph{optimal preview} with the highest score in the
corresponding space of previews. \vspace{-1mm}

\begin{definition}[Optimal Preview Discovery Problem]
The optimization problem of finding an \emph{optimal preview} is defined
as follows, where $\mathbb{P}$ can be any of the aforementioned three
spaces---$\mathbb{P}_{k,n}$, $\mathbb{P}_{k,n,\leq d}$ and $\mathbb{P}_{k,n,\geq d}$.
\begin{equation}\label{eq:opt}
\mathcal{P}_{opt} \in \argmax_{\mathcal{P}\in \mathbb{P}} S(\mathcal{P})
\end{equation}
Note that the $\argmax$ function may return a set of optimal previews
due to ties in scores. \qed \vspace{-1mm}
\end{definition}

For example, given the entity graph in Fig.~\ref{fig:ergraph}, using
coverage-based scoring measures for both key and non-key attributes, an
optimal concise preview consisting of 2 tables and 6 non-key
attributes (i.e., $k$=$2$, $n$=$6$) is $\mathcal{P}=\{T_1:\etype{Film}, \edge{Actor}, \edge{Genres}, \edge{Director}, $ $ \edge{Producer}$; $T_2:\etype{Film Actor}, \edge{Actor}, \edge{Award Winners}\}$.
The edge \edge{Actor} is a non-key attribute in both $T_1$ and $T_2$, in different directions.
An optimal diverse preview under the same size constraint ($k$=$2$, $n$=$6$) and distance constraint $d$=$2$
is $\mathcal{P}=\{T_1:\etype{Film}, \edge{Actor}, \edge{Genres}, \edge{Director},$ $\edge{Producer}, \edge{Executive Producer}; T_2:\etype{Award}, \edge{Award Winners}\}$.

\subsection{\NP-hardness of the Optimal Tight and Diverse Preview Discovery Problems}

The optimal preview discovery problem is non-trivial.
Particularly, the problem in the spaces of both tight previews ($\mathbb{P}_{k,n,\leq d}$) and diverse previews ($\mathbb{P}_{k,n,\geq d}$) is \NP-hard.\vspace{-1mm}

\begin{theorem}\label{theorem:tight-np}
Optimal tight preview discovery is \NP-hard.\vspace{-2mm}
\end{theorem}

\begin{proof}
The decision version of the optimal tight preview discovery problem is $\mathit{TightPreview}(G_s,k,n,d,s)$---Given a schema graph $G_s$, decide whether there exists such a preview $\mathcal{P}$ that (1) $\mathcal{P}$ has $k$ tables and no more than $n$ non-key attributes; (2) the distance between every pair of preview tables is not greater than $d$; and (3) the preview's score is at least $s$, i.e., $S(\mathcal{P}) \geq s$.

We construct a reduction, in polynomial-time, from the \NP-hard Clique problem to $\mathit{TightPreview}(G_s,k,n,d,s)$.  Recall that the decision version of $\mathit{Clique}(G,k)$ is to, given a graph $G(V, E)$, decide whether there exists a clique in $G$ with $k$ vertices.  The reduction is by constructing a schema graph $G_s$ from $G$.  For simplicity of exposition, in both this proof and the proof of Theorem~\ref{theorem:diverse-np}, we assume the schema graph $G_s$ is undirected and every edge $\gamma$ in $G_s$ corresponds to the same relationship type.  This assumption is made without loss of generality.  Note that our following proof casts no requirement on the score of a preview (i.e., $s=0$) and thus no requirement on the scores of key and non-key attributes in $G_s$.  Hence, edge orientation and its corresponding relationship type bears no significance in the proof.

Formally, we construct a schema graph $G_s(V_s$, $E_s)$ from $G$ through a vertex bijection $f : V \rightarrow V_s$:\vspace{-2mm}

\begin{itemize}[leftmargin=*]
  \setlength{\itemsep}{0cm}%
	\item $\forall e(v,v') \in E$, there exists an edge (i.e., relationship type) $\gamma(\tau,\tau')$ $\in$ $E_s$, where $\tau=f(v)$ and $\tau'=f(v')$.\vspace{-1mm}
	\item $\forall \gamma(\tau,\tau') \in E_s$, there exists an edge $e(v,v') \in E$, where $v=f^{-1}(\tau)$ and $v'=f^{-1}(\tau')$.\vspace{-2mm}
\end{itemize}
$Clique(G, k)$ is thus reduced to $\mathit{TightPreview}(G_s,k,k,1,0)$ by the above bijections.\vspace{-2mm}
\end{proof}

The \NP-hardness of the optimal diverse preview discovery problem is also based on a reduction from the Clique problem, although the proof is more complex.\vspace{-2mm}

\begin{theorem}\label{theorem:diverse-np}
Optimal diverse preview discovery is \NP-hard.\vspace{-2mm}
\end{theorem}
\begin{proof}
The decision version of the optimal diverse preview discovery problem is $\mathit{DiversePreview}(G_s,k,n,d,s)$---Given a schema graph $G_s$, decide whether there exists such a preview $\mathcal{P}$ that (1) $\mathcal{P}$ has $k$ tables and no more than $n$ non-key attributes; (2) the distance between every pair of preview tables is not smaller than $d$; and (3) the preview's score is at least $s$, i.e., $S(\mathcal{P}) \geq s$.

We construct a reduction, in polynomial-time, from the \NP-hard $\mathit{Clique}(G,k)$ to $\mathit{DiversePreview}(G_s,k,n,d,s)$.  The reduction is also by constructing a schema graph $G_s($$V_s$, $E_s)$ from $G$.  It is similar to the reduction for $\mathit{TightPreview}(G_s,k,n,d,s)$ in Theorem~\ref{theorem:tight-np}, but also bears two important differences.  (1) $G_s$ contains a special vertex, denoted $\tau_0$, that is directly connected to every other vertex in $G_s$.  (2) Barring $\tau_0$ and all its incident edges, $G_s$ is the complement graph of $G$---There is still a vertex bijection $f : V \rightarrow V_s$, but an edge exists between two vertices in $G_s$ if and only if there is no edge between the corresponding vertices in $G$.
Formally, the construction of $G_s$ from $G$ is as follows:\vspace{-2mm}

\begin{itemize}[leftmargin=*]
  \setlength{\itemsep}{0cm}%
	\item $\forall \tau, \tau' \in V_s \backslash \{\tau_0\}$, $\gamma(\tau,\tau') \in E_s$  if and only if $\nexists e(v,v') \in E$, where $v=f^{-1}(\tau)$ and $v'=f^{-1}(\tau')$.\vspace{-1mm}
	\item $\forall \tau \in V_s \backslash \{\tau_0\}$, $\gamma(\tau_0,\tau) \in E_s$.\vspace{-2mm}
\end{itemize}
$Clique(G, k)$ is thus reduced to $\mathit{DiversePreview}(G_s,k,k,2,0)$ by the above construction of $G_s$.\vspace{-2mm}
\end{proof}

\begin{figure}[t]
\begin{center}
     \vspace{-2mm}
     \includegraphics[width=0.35\textwidth]{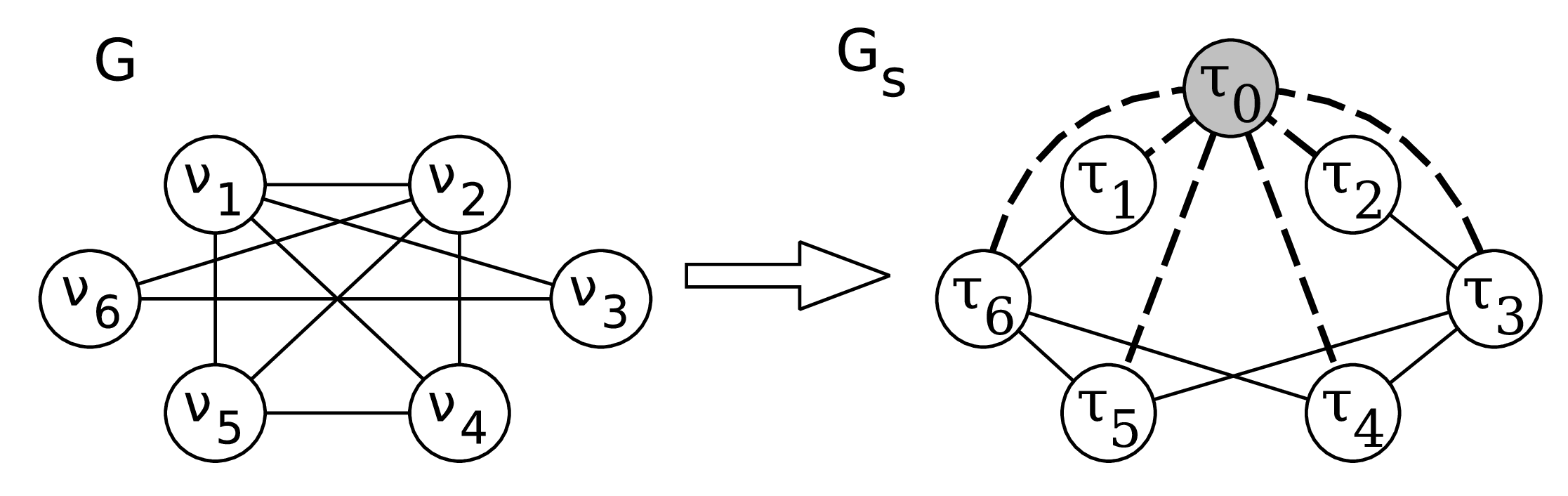}
\end{center} \vspace{-7mm}
\caption{\small Construction of $G_s$ from $G$, for reduction from the clique problem to the optimal diverse preview discovery problem.} \label{fig:npc-proof}
\end{figure}

Fig.~\ref{fig:npc-proof} can help understand the reduction from $Clique(G, k)$ to $\mathit{DiversePreview}$ $(G_s,k,k,2,0)$ in the above proof.
The figure shows an example with $G$ (left) and the constructed schema graph $G_s$ (right), where the gray vertex in $G_s$ is $\tau_0$.  Consider an arbitrary pair of vertices ($v, v'$) in $G$ and their corresponding vertices ($\tau, \tau'$) in $G_s$.  On the one hand, if $v$ and $v'$ are not directly connected in $G$ (e.g., $v_1$ and $v_6$), an edge between $\tau$ and $\tau'$ (i.e., $\tau_1$ and $\tau_6$) is included into $G_s$.  When finding a diverse preview where pairwise table distance must be at least $2$, $\tau$ and $\tau'$ will never be chosen together as the key attributes of two tables in the preview.  Correspondingly, this means a clique must not include both $v$ and $v'$.  On the other hand, if $v$ and $v'$ are directly connected in $G$ (e.g., $v_1$ and $v_2$), there must not be a direct edge between $\tau$ and $\tau'$ (i.e., $\tau_1$ and $\tau_2$) in $G_s$.  The distance between $\tau$ and $\tau'$ is exactly $2$, since they are only indirectly connected through $\tau_0$.  They will thus be considered in choosing the key attributes of two tables in a diverse preview where pairwise table distance must be at least $2$.  Correspondingly, the directly connected $v$ and $v'$ are considered together in forming a clique.

\section{Algorithms}
\label{sec:algorithms}

In this section we discuss algorithms for solving the optimal preview discovery problem.  As given in Eq.~\ref{eq:opt}, the problem is to find a preview with the highest score among candidate previews, where the space of candidates can be concise previews ($\mathbb{P}_{k,n}$), tight previews ($\mathbb{P}_{k,n,\leq d}$) or diverse previews ($\mathbb{P}_{k,n,\geq d}$).   Recall that we use $S(\tau)$ to denote the score of a candidate key attribute $\tau$ for a preview table $T$ and $S^{\tau}(\gamma)$ to denote the score of a candidate non-key attribute $\gamma(\tau, \tau')$ (or $\gamma(\tau', \tau)$) for $T$ whose key attribute is $\tau$.

Our effort focuses on reducing the cost in finding optimal previews.
Both the schema graph and the scoring measures defined in Sec.~\ref{sec:scoring}
are computed before optimal preview discovery.
This is a realistic assumption, since
the schema graph and scoring measures do not change by the size and distance
constraints $k,n,d$.  Furthermore, they can be incrementally updated
when the underlying entity graph is updated (detailed discussion omitted).
On the other hand, the optimal previews cannot be incrementally updated.

Before we present the algorithms, consider the space of all possible previews.
Every entity type $\tau$ can be the key attribute of a preview table $T$.
Let $\Gamma^{\tau}$ denote the set of all edges (i.e., relationship types) incident on $\tau$ in schema graph $G_s$.  Any $\gamma \in \Gamma^{\tau}$ can be a candidate for the non-key attributes of $T$.  By the scoring function in Eq.~\ref{eq:table-score} and the problem formulation in Eq.~\ref{eq:opt}, the non-key attributes of $T$ must have the highest scores among the candidates in $\Gamma^{\tau}$.
This property, stated in Theorem~\ref{theorem:topj}, is important to our algorithms.\vspace{-1mm}

\begin{theorem}\label{theorem:topj}
Suppose an optimal (concise, tight or diverse) preview $\mathcal{P}_{opt}$ contains a preview table $T \in \mathbb{T}$ with key attribute $\tau$.  If $T$ has $m$ non-key attributes, they must be the top-$m$ non-key attributes by scores, i.e., $\forall \gamma, \gamma' \in \Gamma^{\tau}$, if $\gamma \in T.nonkey$ and $\gamma' \notin T.nonkey$, then $S^{\tau}(\gamma) \geq S^{\tau}(\gamma')$.
\end{theorem}

\begin{algorithm}[t]
\scriptsize
\LinesNumbered
\SetKwInOut{Input}{Input}\SetKwInOut{Output}{Output}
\Input{schema graph $G_s$, size constraint $(k, n)$}
\Output{an optimal preview $\mathcal{P}_{opt}$}
\BlankLine

\ForEach{ $\tau \in V_s$ }
{
    $\langle \gamma^{\tau}_1, \gamma^{\tau}_2, \ldots \rangle \leftarrow$ sort the candidate non-key attributes $\gamma^{\tau}_j \in \Gamma^{\tau}$ by their scores $S^{\tau}(\gamma^{\tau}_j)$;\label{line:ordernonkey}
}
$max\_score \leftarrow 0$;
$\mathcal{P}_{opt} \leftarrow \emptyset$;\\
\ForEach{ $k$-\text{subset of} $V_s$ (\text{denoted} $V$)\label{line:enum}}
    {
    $score \leftarrow 0$;
    $\mathcal{P} \leftarrow \emptyset$;
    $i \leftarrow 1$;\label{line:eachsubset}\\
    \ForEach { $\tau \in V$ }
    {
            $\mathcal{P}[i].key = \tau$;\\
            $\mathcal{P}[i].nonkey = \{\gamma^{\tau}_1\}$;\label{line:top1}\\
            $score = score + S(\tau)\times S^{\tau}(\gamma^{\tau}_1)$;\\
            $i \leftarrow i+1$;
    }
    $\Gamma \leftarrow$ top-($n$$-$$k$) candidate non-key attributes from all $\tau \in V$ in descending order of $S(\tau)\times S^{\tau}(\gamma^{\tau}_j)$;\label{line:orderremain}

    \ForEach { $\gamma^{\tau}_j \in \Gamma$, where $\tau=\mathcal{P}[x].key$}
    {
            $score \leftarrow score + S(\tau)\times S^{\tau}(\gamma^{\tau}_j)$;\\
            $\mathcal{P}[x].nonkey \leftarrow \mathcal{P}[x].nonkey \bigcup \{\gamma^{\tau}_j\}$;\label{line:addremain}
    }
    \If{$score > max\_score$}
    {
        	$max\_score \leftarrow score$;\\
    		$\mathcal{P}_{opt} \leftarrow \mathcal{P}$;
    }
}

\Return{$\mathcal{P}_{opt}$;}
\caption{\small Brute-force algorithm for optimal preview discovery}\label{algo:bf}
\end{algorithm}

\vspace{-2mm}
\subsection{A Brute-Force Algorithm}\label{sec:bruteforce}
Alg.~\ref{algo:bf} is a brute-force algorithm for the optimal preview discovery problem.  It enumerates all possible $k$-subsets of entity types, as the $k$ entity types in each subset form the key attributes of $k$ preview tables in a preview $\mathcal{P}$ (Line~\ref{line:enum}).  For a candidate key attribute $\tau$, the elements in the set of its candidate non-key attributes $\Gamma^{\tau}$ are ordered by their scores.  We denote these candidates in descending order of scores by
$\gamma^{\tau}_1$, $\gamma^{\tau}_2$, and so on (Line~\ref{line:ordernonkey}).  Suppose preview table $T$ uses $\tau$ as its key attribute. Each table must contain at least one non-key attribute, according to Definition~\ref{def:preview}.  Hence, $\gamma^{\tau}_1$ (i.e., the candidate non-key attribute with the highest score) must be included into $T.nonkey$ (Line~\ref{line:top1}), by Theorem~\ref{theorem:topj}.  Further, among the remaining candidate non-key attributes for the $k$ entity types, the top-($n$$-$$k$) candidates by scores must be included into $\mathcal{P}$ (Lines~\ref{line:orderremain}--\ref{line:addremain}), by Theorem~\ref{theorem:topj}.  Note that, since the sorted list of candidate non-key attributes for each $\tau$ is already created (Line~\ref{line:ordernonkey}), it is unnecessary to do a full sorting in order to determine the top-($n$$-$$k$) candidates $\Gamma$.  Instead, a simple merge operation on the $k$ sorted lists will get $\Gamma$.

The algorithm has an exponential complexity $O(KN\log N+{K\choose k}(k+n))$, where $K=|V_s|$ is the number of candidate key attributes, $N=2|E_s|$ is the number of candidate non-key attributes for all candidate key attributes, $K\choose k$ is the number of $k$-subsets, and $KN\log N$ is for sorting individual lists of candidates (Line~\ref{line:ordernonkey}), in which each list contains at most $N$ elements.

Alg.~\ref{algo:bf} is for finding one of the optimal previews.  To find all optimal previews, it needs simple extension to deal with ties in scores, which we will not further discuss.

The same brute-force algorithm is applicable for optimal preview discovery in all three types of spaces---concise, tight and diverse previews.  The pseudo code in Alg.~\ref{algo:bf} is for concise previews and does not enforce distance constraint, for simplicity of presentation.  Enforcing distance constraint for tight/diverse previews is straightforward, by performing distance check on every pair of preview tables in each $k$-subset of entity types.

\begin{algorithm}[t]
\scriptsize
\LinesNumbered
\SetKwInOut{Input}{Input}\SetKwInOut{Output}{Output}
\Input{schema graph $G_s$, size constraint $(k, n)$}
\Output{an optimal concise preview $\mathcal{P}_{opt}$}
\BlankLine

\ForEach{$x\leftarrow 1$ \KwTo $K$}
{
    $\langle \gamma^{\tau_x}_1, \gamma^{\tau_x}_2, \ldots \rangle \leftarrow$ sort the candidate non-key attributes $\gamma^{\tau_x}_j \in \Gamma^{\tau_x}$ by their scores $S^{\tau_x}(\gamma^{\tau_x}_j)$;
}
\For{$x\leftarrow 1$ \KwTo $K$}{
    \For{$i\leftarrow 1$ \KwTo $\min(k,x)$}{
        \For{$j\leftarrow i$ \KwTo $n$}{
			$\mathcal{P}_{opt}(i,j,x) \leftarrow \mathcal{P}_{opt}(i,j,x-1)$;\\
            \For{$m\leftarrow 1$ \KwTo $min(j-i+1,|\Gamma^{\tau_x}|)$}{
                $T_x^m.key \leftarrow \tau_x$;\\
                $T_x^m.nonkey \leftarrow$ top-$m$ candidate non-key attributes in $\Gamma^{\tau_x}$;\\ $\mathcal{P} \leftarrow \mathcal{P}_{opt}(i-1,j-m,x-1) \bigcup \{T_x^m\}$;\\
				\If{$S(\mathcal{P}) > S(\mathcal{P}_{opt}(i,j,x))$}
				{
					$\mathcal{P}_{opt}(i,j,x) \leftarrow \mathcal{P}$;\\
				}
			}
		}
	}
}
$\mathcal{P}_{opt} \leftarrow \mathcal{P}_{opt}(k,n,K)$;\\
\Return{$\mathcal{P}_{opt}$;}
\caption{\small Dynamic-programming algorithm for optimal concise preview discovery}\label{algo:dp}
\end{algorithm}

\vspace{-2mm}
\subsection{A Dynamic-Programming Algorithm for Concise Preview Discovery Problem}

As the combinatorial number of $k$-subsets grows exponentially, the performance of the above brute-force algorithm becomes unacceptable for finding an optimal preview under modest size constraints. We thus developed a dynamic-programming algorithm to discover optimal concise previews more efficiently.

Consider an arbitrary order on all $K$ entity types---$\tau_1$, \ldots, $\tau_K$.  We use $\mathcal{P}_{opt}(k,n,x)$ to denote an optimal concise preview among the first $x$ entity types $\tau_1$, \ldots, $\tau_x$.  The optimal concise preview discovery problem is to find $\mathcal{P}_{opt}(k,n,K)$.  $\mathcal{P}_{opt}(k,n,x)$ can be constructed from the solutions to smaller problems, in two ways:  (1) It can be equal to $\mathcal{P}_{opt}(k,n,x$$-$$1)$, i.e., its $k$ tables and $n$ non-key attributes are from the first $x$$-$$1$ entity types and the $x$-th entity type $\tau_x$ does not contribute anything; (2) It can also be the union of $\mathcal{P}_{opt}(k$$-$$1,n$$-$$m,x$$-$$1)$ and a table $T_x^m$, where $\mathcal{P}_{opt}(k$$-$$1,n$$-$$m,x$$-$$1)$ is an optimal preview with $k-1$ tables and $n$$-$$m$ non-key attributes among the first $x$$-$$1$ entity types, and $T_x^m$ is the table whose key attribute is $\tau_x$ and whose non-key attributes are the top-$m$ elements in $\Gamma^{\tau_x}$---the sorted list of candidate non-key attributes for $\tau_x$.  The number $m$ is between $1$ and $n$$-$$(k$$-$$1)$ (or less if there are less than  $n$$-$$(k$$-$$1)$ elements in $\Gamma^{\tau_x}$), since each of the $k$$-$$1$ tables in $\mathcal{P}_{opt}(k$$-$$1,n$$-$$m,x$$-$$1)$ must contribute at least one non-key attribute.  The optimal substructure of the problem is as follows.  (We omit boundary cases ($k=1$ or $x=1$ or $n=k$) for brevity.)
$$
\mathcal{P}_{opt}(k,n,x) = \operatorname*{arg\,max}_{\mathcal{P}\in \mathbb{P}(k,n,x)}S(\mathcal{P})\
$$
\vspace{1mm}
$$
\mathbb{P}(k,n,x) =
\left\{
\begin{array}{l l}
\mathcal{P}_{opt}(k,n,x\!-\!1),\\
\mathcal{P}_{opt}(k\!-\!1,n\!-\!1,x\!-\!1)\, \bigcup\, \{T_x^1\},\\
\mathcal{P}_{opt}(k\!-\!1,n\!-\!2,x\!-\!1)\, \bigcup\, \{T_x^2\},\\
...\\
\mathcal{P}_{opt}(k\!-\!1,k\!-\!1,x\!-\!1)\, \bigcup\, \{T_x^{n-(k-1)}\}
\end{array}
\right\},\vspace{1mm}
$$
where $T_x^m.key$ = $\tau_x$ and $T_x^m.nonkey$ = top-$m$ candidate non-key attributes in $\Gamma^{\tau_x}$.
Note that the optimal substructure is inapplicable when previews must satisfy distance constraint in addition to size constraint (details omitted).  Therefore the dynamic-programming algorithm is for concise previews but not tight/diverse previews.

The pseudo code of the dynamic-programming algorithm is shown in Alg.~\ref{algo:dp}.  Its complexity is $O(KN\log N+Kkn^2)$.  Similar to Alg.~\ref{algo:bf}, Alg.~\ref{algo:dp} is for finding one optimal preview.  Finding all optimal previews requires simple extension to deal with ties in scores, which we will not further discuss.

Both Alg.~\ref{algo:bf} and \ref{algo:dp} assume that, given any $k$ entity types (key attributes), they always together have at least $n$ non-key attributes.  That may not be true in reality.  In fact, for two previews with the same number of tables, the preview with less non-key attributes may have the higher score than the other preview.  Note that, in Eq.~\ref{eq:opt}, the optimal preview is not required to have exactly $n$ non-key attributes.  It is simple to extend Alg.~\ref{algo:bf} and \ref{algo:dp} to fully comply with the definition.  Given any entity type $\tau$, if it has less than $n$ candidate non-key attributes, we can simply pad the sorted list $\Gamma^{\tau}$ by pseudo non-key attributes with zero scores.

\begin{table*}[t]
\noindent \begin{minipage}{0.3\textwidth}
\centering
\scriptsize
\begin{tabular}{|l|l|l|l|}
    \hline
    \textbf{Domain}   & \textbf{\# of vertices} & \textbf{\# of edges} \\ \hline
    books    & 6M / 91     & 15M / 201        \\
    film     & 2M / 63    & 18M / 136       \\
    music    & 27M / 69   & 187M / 176    \\
    TV       & 2M / 59   & 17M / 177   \\
    people   & 3M / 45   & 17M / 78     \\
    basketball & 19K / 6  & 557K / 21 \\
    architecture & 133K / 23 & 432K / 48 \\
    \hline
\end{tabular}
\vspace{-3mm}\caption{\small Sizes of entity/schema graphs.}\vspace{-4mm}
\label{tab:dataset-freebase}
\end{minipage}
\hspace{2mm}
\noindent \begin{minipage}{0.25\textwidth}
\centering
\scriptsize
\begin{tabular}{|l|l|l|l|}
\hline
    \textbf{Domain}   & \textbf{Coverage} & \textbf{Entropy} \\ \hline
    books    & 0.8      & 0.786         \\
    film     & 0.2      & 0.25        \\
    music    & 0.528    & 0.589      \\
    TV       & 0.622    & 0.379    \\
    people   & 0.708    & 0.606      \\
\hline
\end{tabular}
\vspace{-1mm}\caption{\small MRR of non-key attribute scoring.}\vspace{-3mm}
\label{tab:attr-mrr}
\end{minipage}
\hspace{2mm}
\noindent \begin{minipage}{0.4\textwidth}
\centering
\scriptsize
\begin{tabular}{|p{0.8 cm}|p{0.8 cm}|p{0.8 cm}|p{0.8 cm}|p{0.8 cm}|p{0.8 cm}|}
\hline
    & \multicolumn{3}{ c| }{\textbf{key attribute}} & \multicolumn{2}{ c| }{\textbf{non-key attribute}} \\ \hline
    \textbf{Domain}   & \textbf{YPS09} & \textbf{Coverage} & \textbf{Random Walk} & \textbf{Coverage} & \textbf{Entropy} \\ \hline
    books    & 0.4 & 0.55           & 0.43    &   0.43           & 0.43      \\
    film     & -0.01 & 0.48           & 0.25    &   0.35           & 0.35      \\
    music  & 0.37 & 0.33           & 0.46      &   0.42           & 0.41     \\
    TV    & 0.37 & 0.69           & 0.65       &  0.47           & 0.47       \\
    people & 0.36 & 0.31           & 0.29      &    0.43           & 0.43     \\
\hline
\end{tabular}
\vspace{-3mm}\caption{\small PCC of key and non-key attribute scoring.}\vspace{-4mm}
\label{tab:pcc}
\end{minipage}
\end{table*}

\begin{algorithm}[t]
\scriptsize
\LinesNumbered
\SetKwInOut{Input}{Input}\SetKwInOut{Output}{Output}
\Input{schema graph $G_s$, size constraint$(k, n)$, distance constraint $d$}
\Output{an optimal tight/diverse preview $\mathcal{P}_{opt}$}
\BlankLine

$\mathcal{L}_2 \leftarrow \varnothing$\;\label{line:begin-step1}
\ForEach{$i\leftarrow 1$ \KwTo $K$}
{
    \ForEach{$j\leftarrow i+1$ \KwTo $K$}
    {
        \If(\tcc*[f]{$\geq d$ \textrm{for diverse preview}}){$dist(\tau_i,\tau_j) \leq d$}
        {
            $\mathcal{L}_2 \leftarrow \mathcal{L}_2 \cup \{ \langle i\ j\rangle \}$\;
        }
    }
}

$i \leftarrow 3$\;
\While {$i \leq k$ \emph{and} $\mathcal{L}_{i-1} \neq \varnothing$}
{
	$\mathcal{L}_i \leftarrow \varnothing$\;
	\ForEach{$A,B\in \mathcal{L}_{i-1}$ s.t. $(\forall j < i-1: A[j] = B[j])$ \emph{and} $(A[i-1]<B[i-1])$}
	{
        \tcc{$\geq d$ \textrm{for diverse preview}}
		\If{$dist(\tau_{A[i-1]},\tau_{B[i-1]}) \leq d$ \label{line:checkdistance}}
		{
			$\mathcal{L}_i \leftarrow \mathcal{L}_i \cup \{  \langle A[1] \ldots A[i-1]\ B[i-1]\rangle \}$\;
		}
	}
	$i\leftarrow i+1$\;
}
\If{$\mathcal{L}_k = \varnothing$}
{
	\Return $\varnothing$\;\label{line:end-step1}
}
$max\_score \leftarrow 0$;\label{line:begin-step2}\\
\ForEach{$A \in \mathcal{L}_k$}
{
	$\mathcal{P}\leftarrow \mathit{ComputePreview}(A)$\;\label{line:comppreview}
	\If{$score(\mathcal{P})>max\_score$}
	{
        	$max\_score \leftarrow score(\mathcal{P})$;\\
    		$\mathcal{P}_{opt} \leftarrow \mathcal{P}$;\label{line:end-step2}
	}
}
\Return $\mathcal{P}_{opt}$;
\caption{\small Apriori-style Algorithm for optimal tight/diverse preview discovery}\label{algo:tightpreview}
\end{algorithm}

\vspace{-2mm}
\subsection{An Apriori-style Algorithm for Tight / Diverse Preview Discovery Problem}

Since the dynamic-programming algorithm is inapplicable when previews must satisfy distance constraint, we propose an efficient algorithm for optimal tight/diverse preview discovery, shown in Alg.~\ref{algo:tightpreview}. It consists of two steps: (1) finding $k$-subsets of entity types (i.e., vertices in $G_s$) satisfying the distance constraint (Lines~\ref{line:begin-step1}--~\ref{line:end-step1}); (2) for each qualifying $k$-subset of entity types, forming a preview under the size constraint, computing its score and choosing a preview with the highest score (Lines~\ref{line:begin-step2}--~\ref{line:end-step2}).

The first step is essentially finding $k$-cliques in a graph converted from the schema graph $G_s$, in which vertices are considered adjacent if they are within distance $d$ (for tight previews) or apart by at least distance $d$ (for diverse previews).  The $k$-clique problem is well-studied and many efficient algorithms have been designed in the past.  Our method is inspired by the well-known Apriori algorithm~\cite{apriori} for frequent itemset mining.  In~\cite{journals/bioinformatics/KoseWLF01}, an algorithm was proposed for finding $k$-cliques (where edges correspond to metabolite correlations) by similar ideas, although the connection to Apriori was not made.  Their experimental results demonstrated superior efficiency in comparison with the more well-known Bron-Kerbosch algorithm~\cite{Bron-Kerbosch}.  Nevertheless, the two broad steps of our optimal tight/diverse preview discovery algorithm are independent from each other, and thus any more efficient or even approximate algorithm for finding $k$-cliques can be plugged into it to further improve its execution efficiency.

In more details, the first step of Alg.~\ref{algo:tightpreview} iteratively generates a $k$-subset of entity types by merging two $(k$$-$$1)$-subsets.  Entity types are arbitrarily ordered as $\tau_1, \ldots, \tau_K$. In the $i$-th iteration of the algorithm, if two $(i$$-$$1)$-subsets $A$ and $B$ only differ by their last entity types $\tau_{A[i-1]}$ and $\tau_{B[i-1]}$, and the distance between their last entity types satisfies the distance constraint, a candidate $i$-subset is generated by appending $\tau_{B[i-1]}$ to the end of $A$.

In the second step, for each candidate $k$-subset of entity types, a preview is computed ($\mathit{ComputePreview}(A)$ in Line~\ref{line:comppreview} of Alg.~\ref{algo:tightpreview}). The details of function $\mathit{ComputePreview}$ are omitted.  It follows Theorem~\ref{theorem:topj} and is essentially the same as Lines~\ref{line:eachsubset}--~\ref{line:addremain} in Alg.~\ref{algo:bf}.
The score of each preview is computed (the same as in Lines~\ref{line:eachsubset}--~\ref{line:addremain} of Alg.~\ref{algo:bf}) and a preview with the highest score is returned.

The worst-case complexity of Alg.~\ref{algo:tightpreview} is the same as that of Alg.~\ref{algo:bf}.  However, as Sec.~\ref{sec:experiments} shows, in practice it significantly outperforms the brute-force algorithm, since Line~\ref{line:checkdistance} could filter out many combinations that do not satisfy the distance constraint.

\section{Evaluation}
\label{sec:experiments}

We conducted experiments to evaluate the preview scoring
measures' accuracy (Sec.~\ref{sec:exp-score}), the preview discovery
algorithms' efficiency (Sec.~\ref{sec:exp-eff}), and the overall
quality of discovered previews (Sec.~\ref{sec:exp-sample}).
All experiments were run on a Dell T100 server running Ubuntu 8.10.
The server has a Dual Core Xeon E3120 processor, 6MB cache, 4GB RAM,
and two 250GB RAID1 SATA hard drivers.  All algorithms are implemented in
C++ and compiled with `-O2' optimization in GCC-4.3.2.

The entity graph used in our experiment is a dump of Freebase at
September 28, 2012.\footnote{\small \url{https://developers.google.com/freebase/data}}
The dataset is imported into an MySQL database.
In Freebase, the entire entity graph is partitioned into many domains.
Our experiments were conducted on seven domains.  The sizes of
the entity and schema graphs in these domains are shown in Table~\ref{tab:dataset-freebase}.
Our work currently is limited to named entities, thus all numeric attribute values from the data dump have been removed.  Note that a schema graph may be disconnected.  To ensure the convergence of random walk in such a graph, we added a small transition probability $10^{-5}$ to every pair of entity types.

\vspace{-1mm}
\subsection{Accuracy of Preview Scoring Measures}
\label{sec:exp-score}

We conducted two experiments to evaluate the accuracy of the scoring measures for both key and non-key attributes presented in Sec.~\ref{sec:scoring}.  One experiment compares the ranking orders of candidate key (non-key) attributes by the scoring measures with gold standard ranking orders.  The other calculates the correlation between two pairwise ordering results on candidate key (non-key) attributes---one by the scoring measures and the other collected through crowdsourcing. In both experiments, we used both measures proposed in this paper and an adaptation of the approach in~\cite{DBLP:journals/pvldb/YangPS09}.\vspace{-1mm}

\subsubsection{Adaptation of~\cite{DBLP:journals/pvldb/YangPS09}}\label{sec:yps09}

Yang et al.~\cite{DBLP:journals/pvldb/YangPS09} proposed an algorithm to summarize relational databases, specifically the tables in TPC-E benchmark.~\footnote{\small \url{http://www.tpc.org/tpce/}} Their approach works in three steps. First they define an importance value for each table considering both information content of the tables and join relationships between the tables. Second, they measure the similarity/distance between tables.  Finally, they use a weighted $k$-center clustering algorithm to place the tables into $k$ clusters. The $k$ cluster centers are the summary of the database. We implemented their algorithm. We compared the results on TPC-E tables with those reported in~\cite{DBLP:journals/pvldb/YangPS09} and validated our own implementation.

We adapted~\cite{DBLP:journals/pvldb/YangPS09} for the entity graphs in the aforementioned Freebase domains. Since~\cite{DBLP:journals/pvldb/YangPS09} was designed to summarize relational databases only, we converted each entity graph into a relational database, as follows. For each entity type $\tau$, we created a relational table, of which the first column takes entities belonging to $\tau$ as its values. Furthermore, a column is created for each relationship type incident on $\tau$ in the scheme graph. The values in such a column are the entities adjacent to the entities in the first column through the corresponding relationship type.
For each entity belonging to $\tau$, a number of tuples are inserted into the table, which are essentially a Cartesian product of distinct values on all these columns.\vspace{-1mm}

\subsubsection{Comparison with Gold Standard}\label{sec:gold}
\vspace{-2mm}{\flushleft \textbf{Key attributes}:}

We collected gold standard data for 5 largest entity domains in Freebase---``books'', ``film'', ``music'', ``TV'' and ``people''.  For each domain, Freebase offers an entrance page showing 6 major entity types in that domain.  A user can choose to browse entities in any of the 6 types.~\footnote{\small The entrance pages were all under ``Featured Data'' on \url{Freebase.com}.  For instance, \url{http://www.freebase.com/view/film} was the entrance page for domain ``film''.  We collected these pages shortly after September 28, 2012, which is the timestamp of the Freebase entity graph dump used in our experiments.
These pages have become unavailable lately.}  As such entrance pages were manually created by Freebase, our conjecture is that they are of high quality and reflect the most popular entity types.  We thus treated the 6 entity types listed in the entrance page of a domain as the gold standard for top-6 key attributes in that domain.
The schema of the tables in the gold standard can be found in Table~\ref{tab:appendix-gold-standard} in the Appendix.

\begin{figure}[t]
\begin{center}
     \includegraphics[width=0.48\textwidth]{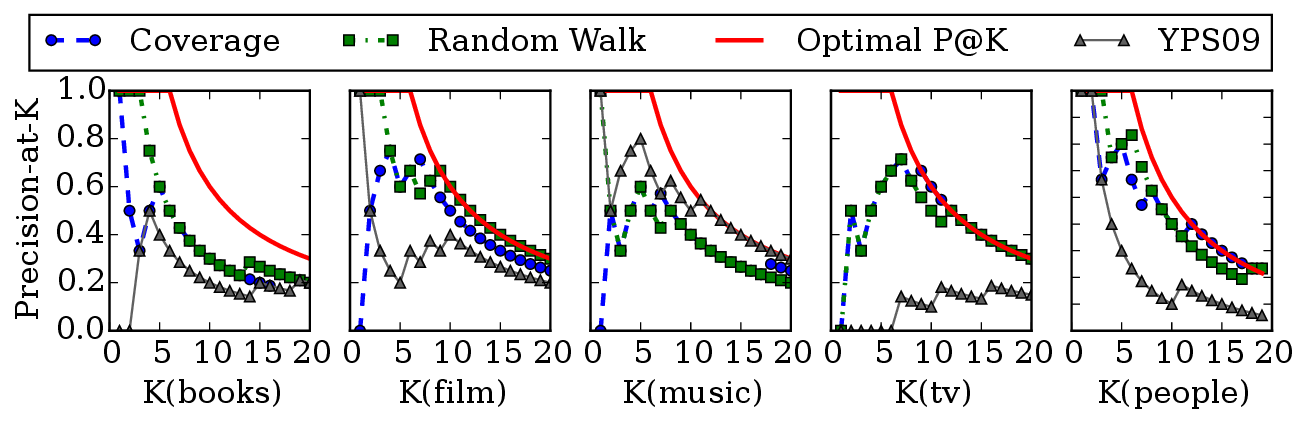}
\end{center} \vspace{-8mm}
\caption{\small Precision-at-$K$ of key attribute scoring.} \label{fig:ctype-util}\vspace{-3mm}
\end{figure}

For both the coverage-based and the random-walk based scoring measures in Sec.~\ref{sec:key-score}, we ranked all candidate key attributes by their scores.
We calculated the accuracy of a scoring measure by several widely-used measures, including Precision-at-$K$ (P@$K$), Average Precision (AvgP) and Normalized Discounted Cumulative Gain (nDCG)~\cite{Manning08}.
An approach that ranks accurate results higher is expected to receive better values under these measures.
For a scoring measure for key attributes, P@$K$ is the percentage of its top-$K$ results that belong to the aforementioned gold standard top-6 key attributes.
For the adaptation of~\cite{DBLP:journals/pvldb/YangPS09}, we use the ranked list by their table importance scoring.
The results are in Fig.~\ref{fig:ctype-util}.
The topmost curves (``Optimal P@$K$'') represent the best possible P@$K$ that can be archived by any method.
For instance, P@10 can be at most 0.6, since there are only 6 gold standard key attributes in each domain, as mentioned above.
Fig.~\ref{fig:ctype-util} shows that both the coverage-based and the random-walk based scoring measures had
P@10 close to 0.6 in 4 out of the 5 domains.
They both had significantly higher P@K values than \cite{DBLP:journals/pvldb/YangPS09} (denoted ``YSP09'')
in 4 out of the 5 domains and similar values in the remaining domain.

\begin{figure}[t]
\begin{center}
     \includegraphics[width=0.48\textwidth]{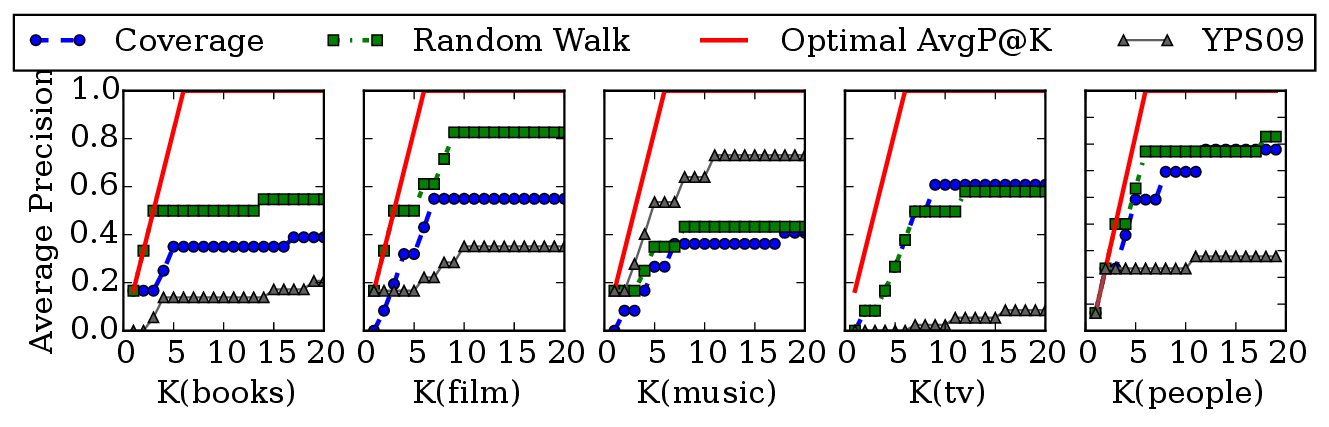}
\end{center} \vspace{-8mm}
\caption{\small Average precision of key attribute scoring.} \label{fig:ctype-map}\vspace{-1mm}
\end{figure}

\begin{figure}[t]
\begin{center}
     \includegraphics[width=0.48\textwidth]{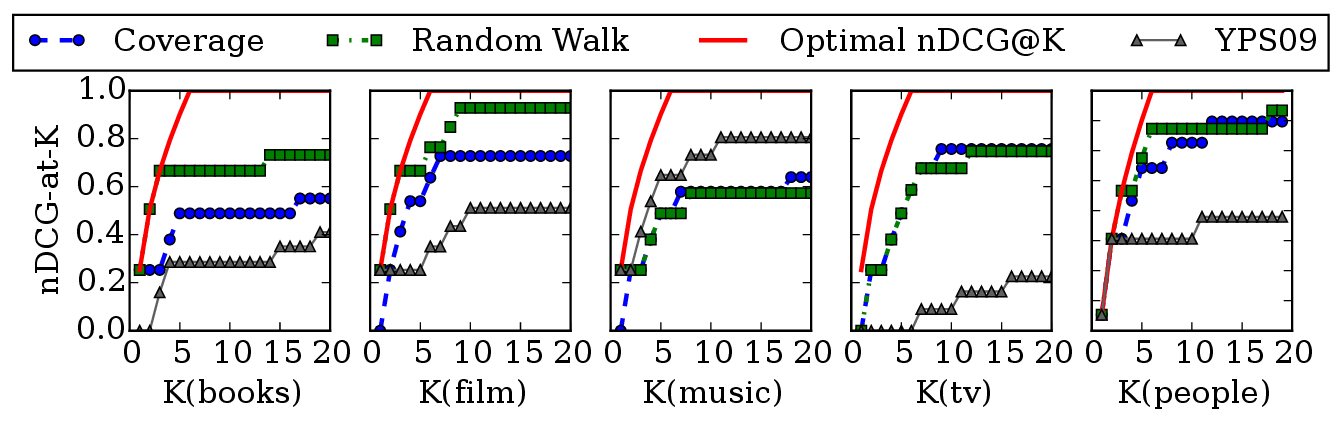}
\end{center} \vspace{-8mm}
\caption{\small nDCG of key attribute scoring.} \label{fig:ctype-ndcg}\vspace{-1mm}
\end{figure}

We also used AvgP and nDCG to gauge the accuracy of the scoring measures for key attributes. The results are as follows:\vspace{-1mm}
\begin{list}{$\bullet$}
{ \setlength{\leftmargin}{1em} \setlength{\itemsep}{-3pt} }

\item Average Precision (AvgP): The average precision of the top-$k$
results is given by AvgP$=\frac{\sum_{i=1}^k{\text{P}@i}\ \times\ rel_i}{\text{size of ground truth}}$,
where $rel_i$ equals $1$ if the result at rank $i$ is in the ground truth
and $0$ otherwise.  Fig.~\ref{fig:ctype-map} shows significantly higher AvgP for both the coverage-based and the random-walk based scoring measures, compared to ~\cite{DBLP:journals/pvldb/YangPS09}, in 4 out of 5 domains.

\item Normalized Discounted Cumulative Gain (nDCG):
The cumulative gain of the top-$k$ results is
DCG$_k$$=$$rel_1$$+$$\sum_{i=2}^{k}{\frac{rel_i}{\log_2(i)}}$.
It penalizes the results if a ground truth result is ranked low.
DCG$_k$ is normalized by IDCG$_k$, the cumulative gain for an ideal ranking
of the top-$k$ results. Thus nDCG$_k$$=$$\frac{\text{DCG}_k}{\text{IDCG}_k}$.
It is shown in Fig.~\ref{fig:ctype-ndcg} that both the coverage-based and the random-walk based scoring measures had clearly higher nDCG, in comparison with ~\cite{DBLP:journals/pvldb/YangPS09}, in 4 out of the 5 domains.
\end{list}

{\flushleft \textbf{Non-key attributes}:}

For each entity type, Freebase offers a table for users to browse and query the entities belonging to that type.~\footnote{\small \url{http://www.freebase.com/music/artist?instances=}, for instance, would display a table for type \etype{Artist} in ``music'' domain.}  The table always has 3 common columns for recording names, types and article contents of entities.  It also has 3 or less type-dependent non-key attributes manually selected by Freebase editors.  Although Freebase allows users to add more attributes into this table, we believe the original $3$ type-dependent attributes in general bear higher quality.  We thus treated these attributes as the gold standard for top non-key attributes for that entity type.

For both the coverage-based and the entropy-based scoring measures in Sec.~\ref{sec:non-key-score}, we ranked all candidate non-key attributes by their scores.
There is no comparison with~\cite{DBLP:journals/pvldb/YangPS09} regarding non-key attributes, since it does not have an component that can be adapted for discovering non-key attributes.
We calculated the accuracy of a scoring measure by Mean Reciprocal Rank (MRR)~\cite{Manning08} instead of P@$K$ as there are only 3 or less gold standard answers for top non-key attributes in each entity type.  For a scoring measure for non-key attributes, the reciprocal rank is the multiplicative inverse of the rank of the first gold standard non-key attribute among its ranking results.  MRR is the average reciprocal rank across all entity types with at least 5 candidate non-key attributes.  (If an entity type has only less than 5 candidates, the gold standard answers are ranked deceptively high.  Thus we exclude such entity types, to obtain more accurate evaluation.)
The results are shown in Table~\ref{tab:attr-mrr}.  In every domain except ``film'' and for both the coverage-based and the entropy-based measures, MRR is above 0.5.
This means in average a gold standard non-key attribute appeared in the top-2 ranked results.
The lower MRR for ``film'' domain is from only one entity type and thus is not truly indicative, since only that entity type has at least
5 candidate non-key attributes.

\vspace{-2mm}
\subsubsection{Correlation with Crowd Ranking}
We conducted an extensive study in Amazon Mechanical Turk (AMT)---a popular crowdsourcing service---and measured the correlation between our scoring measures and users' opinions with regard to key and non-key attributes ranking.  We explain the procedure for evaluating key attribute ranking in one domain, since the procedure is repeated for all 5 gold standard domains and is the same for both key and non-key attribute ranking.

Given a domain, we randomly generated $50$ pairs of entity types, i.e., candidate key attributes.  Each pair was presented to $20$ AMT workers.  The workers were asked which of the 2 entity types in the pair is more important.  To help them understand the tasks, we provided a few examples to explain what are considered more important in common sense.  The workers were also asked to answer a few screening questions that test their common knowledge. They must answer the screening questions correctly, otherwise their responses are not considered.

We collected $1,000$ opinions ($50$ pairs $\times$ $20$ workers per pair) in total.  We then constructed two lists---$X$ and $Y$, each of which contains $50$ values corresponding to the $50$ pairs.  A value in $X$ represents the difference in the ranking positions (by our scoring measures, or by the table importance measure in~\cite{DBLP:journals/pvldb/YangPS09}) of the two entity types in the corresponding pair.  A value in $Y$ represents the difference in the numbers of AMT workers favoring the two entity types.  The correlation between $X$ and $Y$ is measured by Pearson Correlation Coefficient (PCC)~\cite{pcc_cohen} as follows.\vspace{-2mm}

\begin{align}
PCC =  \frac{\text{E}(XY)-\text{E}(X)\text{E}(Y)}{\sqrt{\text{E}(X^2)-(\text{E}(X))^2}\sqrt{\text{E}(Y^2)-(\text{E}(Y))^2}}\vspace{1mm}
\end{align}

The PCC value ranging from $-1$ to $1$ indicates the degree of correlation between the pairwise ranking orders produced by our scoring methods and the pairwise preferences given by AMT workers.
A PCC value in the ranges of [$0.5$,$1.0$], [$0.3$,$0.5$) and [$0.1$,$0.3$) indicates a strong, medium and small positive correlation, respectively.
PCC values for the 5 gold standard domains are in Table~\ref{tab:pcc}.
For all 5 domains, the results show at least a medium positive correlation between our scoring measures and AMT workers.
For 4 out of the 5 domains, the coverage-based and/or the random walk-based measures had significantly higher PCC values than the
adaptation of~\cite{DBLP:journals/pvldb/YangPS09}(``YPS09''), which even demonstrated slightly negative correlation in the ``film'' domain.

\vspace{-1mm}
\subsection{Efficiency of Algorithms}
\label{sec:exp-eff}

\usetikzlibrary{patterns}

\begin{figure}[t]\centering
\ref{named}
\begin{tikzpicture}
\pgfplotsset{
footnotesize,
height=2.8cm,
width=3.5cm,
xtick=data,
ymin=1,
ymax=10000000,
enlarge x limits=0.2,
xlabel style={yshift=0.5em},
ylabel style={yshift=-1em},
tick label style={font=\scriptsize },
label style={font=\scriptsize},
title style={font=\scriptsize},
/pgfplots/bar  cycle  list/.style={/pgfplots/cycle  list={%
{blue,fill=blue!30!white,mark=none,postaction={pattern=north east lines}},
{red,fill=red!30!white,mark=none},%
}
},
}
\matrix[column sep=0.05cm]
{
\begin{semilogyaxis}[
title={$k$=$5$,$n$=$10$},
xlabel={domain},
ybar,
bar width=4pt,
ylabel={Execution Time (ms)},
xticklabels={B,A,M},
xtick={1,2,3},
legend columns=-1,
legend entries={Brute-Force Algorithm, Dynamic-Programming Algorithm},
legend to name=named,
legend style={font=\scriptsize},
]
\addplot coordinates
{(1, 2.000 )
(2, 60.000 )
(3, 2700.000 )
};
\addplot coordinates
{(1, 3.333 )
(2, 6.667 )
(3, 6.667 )
};
\end{semilogyaxis}
&
\begin{semilogyaxis}[
title={music,$n$=$20$},
xlabel={$k$},
ybar,
bar width=4pt,
yticklabels={,,},
]
\addplot coordinates
{( 3 , 30.000 )
( 6 , 23396.667 )
( 9 , 3433603.333 )
};
\addplot coordinates
{( 3 , 16.667 )
( 6 , 20.000 )
( 9 , 30.000 )
};
\end{semilogyaxis}
&
\begin{semilogyaxis}[
title={music,$k$=$6$},
xlabel={$n$},
ybar,
bar width=4pt,
yticklabels={,,},
]
\addplot coordinates
{( 8 , 18103.333 )
( 12 , 19746.667 )
( 16 , 21840.000 )
( 20 , 23396.667 )
};
\addplot coordinates
{( 8 , 10.000 )
( 12 , 13.333 )
( 16 , 20.000 )
( 20 , 20.000 )
};
\end{semilogyaxis}
\\
};
\end{tikzpicture}
\vspace{-6mm}
\caption{\small Execution time of optimal concise preview discovery algorithms.}\vspace{-4mm}
\label{fig:exp-perf1}
\end{figure}
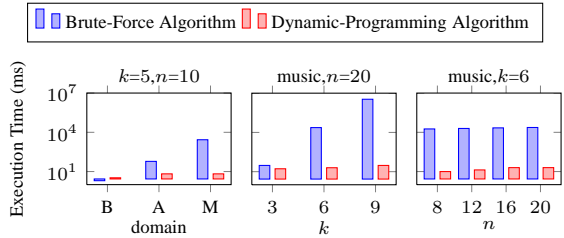

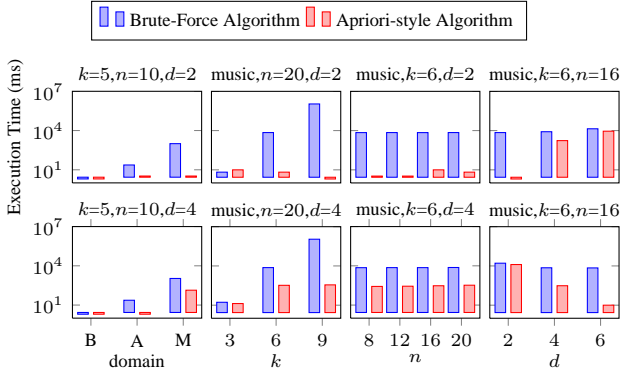
\begin{figure}\centering
\ref{named2}
\begin{tikzpicture}
\pgfplotsset{
footnotesize,
height=2.8cm,
width=3.3cm,
xtick=data,
ymin=1,
ymax=10000000,
enlarge x limits=0.2,
xlabel style={yshift=0.5em},
ylabel style={yshift=-1.5em},
tick label style={font=\scriptsize },
label style={font=\scriptsize},
title style={font=\scriptsize},
/pgfplots/bar  cycle  list/.style={/pgfplots/cycle  list={%
{blue,fill=blue!30!white,mark=none,postaction={pattern=north east lines}},%
{red,fill=red!30!white,mark=none},
}
},
}
\matrix[column sep={1.85cm,between origins}, row sep={1.8cm,between origins}]
{
\begin{semilogyaxis}[
title={$k$=$5$,$n$=$10$,$d$=$2$},
ybar,
bar width=4pt,
ylabel={Execution Time (ms)},
xticklabels={B,A,M},
xtick=\empty,
legend columns=-1,
legend entries={Brute-Force Algorithm, Apriori-style Algorithm},
legend to name=named2,
legend style={font=\scriptsize},
]
\addplot coordinates
{(1, 2.000 )
(2, 23.333 )
(3, 996.667 )
};
\addplot coordinates
{(1, 2.000 )
(2, 3.333 )
(3, 3.333 )
};
\end{semilogyaxis}
&
\begin{semilogyaxis}[
title={music,$n$=$20$,$d$=$2$},
ybar,
bar width=4pt,
xtick=\empty,
yticklabels={,,},
]
\addplot coordinates
{( 3 , 6.667 )
( 6 , 7076.667 )
( 9 , 1086713.333 )
};
\addplot coordinates
{( 3 , 10.000 )
( 6 , 6.667 )
( 9 , 2.000 )
};
\end{semilogyaxis}
&
\begin{semilogyaxis}[
title={music,$k$=$6$,$d$=$2$},
ybar,
bar width=4pt,
xtick=\empty,
yticklabels={,,},
]
\addplot coordinates
{( 8 , 7023.333 )
( 12 , 7060.000 )
( 16 , 7016.667 )
( 20 , 7076.667 )
};
\addplot coordinates
{( 8 , 3.333 )
( 12 , 3.333 )
( 16 , 10.000 )
( 20 , 6.667 )
};
\end{semilogyaxis}
&
\begin{semilogyaxis}[
title={music,$k$=$6$,$n$=$16$},
ybar,
bar width=4pt,
xtick=\empty,
yticklabels={,,},
]
\addplot coordinates
{( 2 , 7056.667 )
( 4 , 8203.333 )
( 6 , 13763.333 )
};
\addplot coordinates
{( 2 , 2.000 )
( 4 , 1716.667 )
( 6 , 9000.000 )
};
\end{semilogyaxis}
\\
\begin{semilogyaxis}[
title={$k$=$5$,$n$=$10$,$d$=$4$},
ybar,
bar width=4pt,
xlabel={domain},
xticklabels={B,A,M},
xtick={1,2,3},
legend columns=-1,
legend entries={Brute-Force Algorithm, Apriori-style Algorithm},
legend to name=named2,
legend style={font=\scriptsize},
]
\addplot coordinates
{(1,2.000)
(2,23.333)
(3,1080.000)
};
\addplot coordinates
{(1,2.000)
(2,2.000)
(3,136.667)
};
\end{semilogyaxis}
&
\begin{semilogyaxis}[
title={music,$n$=$20$,$d$=$4$},
ybar,
bar width=4pt,
yticklabels={,,},
xlabel={$k$},
]
\addplot coordinates
{( 3 , 16.667 )
( 6 , 7456.667 )
( 9 , 1101786.667 )
};
\addplot coordinates
{( 3 , 13.333 )
( 6 , 330.000 )
( 9 , 356.667 )
};
\end{semilogyaxis}
&
\begin{semilogyaxis}[
title={music,$k$=$6$,$d$=$4$},
ybar,
bar width=4pt,
yticklabels={,,},
xlabel={$n$},
]
\addplot coordinates
{( 8 , 7223.333 )
( 12 , 7296.667 )
( 16 , 7283.333 )
( 20 , 7456.667 )
};
\addplot coordinates
{( 8 , 266.667 )
( 12 , 276.667 )
( 16 , 300.000 )
( 20 , 330.000 )
};
\end{semilogyaxis}
&
\begin{semilogyaxis}[
title={music,$k$=$6$,$n$=$16$},
ybar,
bar width=4pt,
yticklabels={,,},
xlabel={$d$},
]
\addplot coordinates
{( 2 , 16176.667 )
( 4 , 7226.667 )
( 6 , 7050.000 )
};
\addplot coordinates
{( 2 , 12780.000 )
( 4 , 306.667 )
( 6 , 10.000 )
};
\end{semilogyaxis}
\\
};
\end{tikzpicture}
\vspace{-1cm}
\caption{\small Execution time of optimal tight (upper) and diverse (lower) preview discovery algorithms.}
\label{fig:exp-perf2}
\end{figure}

This section presents results on the efficiency of the optimal preview discovery algorithms in Sec.~\ref{sec:algorithms}.
On optimal concise preview discovery, we compared the Brute-Force Alg.~\ref{algo:bf} and the Dynamic-Programming Alg.~\ref{algo:dp}.  Specifically, we compared their execution times by varying: (1) size of schema graph (i.e., number of candidate key attributes ($K$) and number of candidate non-key attributes ($N$)); (2) number of preview tables (i.e., key attributes) in a preview ($k$); and (3) maximum number of non-key attributes in a preview ($n$).  For (1), we fixed $k$=$5$, $n$=$10$ and experimented with $3$ domains---``basketball'' (B), ``architecture'' (A), and ``music'' (M). They differ greatly in the sizes of their schema graphs (B: $K$=$6$, $N$=$21$; A: $K$=$23$, $N$=$48$; M: $K$=$69$, $N$=$176$).
For (2), we varied $k$ from $3$ to $9$, fixed $n$=$20$ and used ``music'' domain.
For (3), we varied $n$ from $8$ to $20$, fixed $k$=$6$ and used ``music'' domain.

On optimal tight/diverse preview discovery, we compared the Brute-Force Alg.~\ref{algo:bf} and the Apriori-style Alg.~\ref{algo:tightpreview}, by varying not only the aforementioned 3 parameters but also the distance constraint on $d$.  When we varied other parameters, $d$ is fixed at $2$ and $4$ for tight and diverse previews, respectively.  When we fixed other parameters, $d$ was varied from $2$ to $6$.

The results are in Figs.~\ref{fig:exp-perf1} and \ref{fig:exp-perf2}. In all results, the execution time is averaged across 3 runs, and execution time less than 1 millisecond is rounded to 1 millisecond.
The results show that both the Dynamic-Programming and the Apriori-style algorithms outperformed the Brute-Force algorithm by orders of magnitude in most cases.  The exceptions are the smallest domain ``basketball'' and when the number of requested preview tables is small ($k$=$3$).
In these cases, the overheads of complex data structures and calculations in the advanced algorithms outweighed their benefits.

Fig.~\ref{fig:exp-perf2} shows that the Apriori-style algorithm did not perform well for $d$=$6$ in tight preview discovery and $d$=$2$ in diverse preview discovery.  It is due to the excessive number of candidate $k$-subsets that satisfy the distance constraint in such cases.  For instance, the diameter of a schema graph typically is not large.  In the schema graph of ``film'' domain, the longest path length is 7 and the average path length is around 3--4.  Setting distance constraint $d$=$6$ in finding tight previews will make most previews ``tight''.  It is unnecessary to enforce such a distance constraint.

\vspace{-1mm}
\subsection{User Study}
We conducted an extensive user study to compare seven different approaches, including concise previews (``Concise''), tight previews (``Tight''), diverse previews (``Diverse''), Freebase gold standard (``Freebase'', cf. Sec.~\ref{sec:gold} and Table~\ref{tab:appendix-gold-standard} in Appendix), hand-crafted previews by experts (``Experts''), schema summarization based on \cite{DBLP:journals/pvldb/YangPS09} (``YPS09''), and directly using schema graphs (``Graph'').  For each approach, we created a website for presenting schema information using the approach and collecting participants' responses, on the five domains---``books'', ``film'', ``music'', ``TV'', and ``people''.

To produce hand-crafted previews, we used a group of $10$ experts (Ph.D. students in the database area at the authors' institution).
Each expert participant was rewarded a \$$20$ gift card.
For each domain, we set the expected numbers of key attributes ($K$) and non-key attributes ($N$) to be the same as the values in the Freebase gold standard.
Each expert was requested to produce preview tables under the size constraints given by $K$ and $N$.
During the process, the experts had access to the Freebase website, to help them understand the data.
After the experts worked on all the five domains, they were asked to discuss and submit one consolidated preview for each domain.
We use the consolidated previews as the hand-crafted previews in the ensuing user study.
On average an expert spent about $10$ minutes on the simplest domain ``people'' and more than $30$ minutes on the most complex domain ``film''.
After that, the experts spent about $2$ hours to discuss.
The preview tables from the experts have a reasonable overlap with the ``Freebase'' gold standard, but they also differ substantially, as shown in Tables~\ref{tab:precision-at-k-gold} and \ref{tab:precision-at-k-experts} in Appendix.
The substantial amount of time spent by the experts individually and as a group suggests that it is a challenging and time-consuming process to generate preview tables.
This motivates the need for an automatic approach.

The participants of the user study include $84$ computer science graduate students in the authors' institution.
They all have taken database courses.
None of them was affiliated with the authors' research group or exposed to the research project.
Each participant was rewarded a \$$15$ gift card.

Each participant was randomly assigned to use one of the aforementioned seven approaches (websites).
Each approach received $10$ to $13$ participants.
Before a participant started their session, they were given a $20$-minute introduction on the approach of presenting schema information that they are using.
The participant used the assigned approach to work on all five domains, in the order of ``books'', ``film'', ``music'', ``TV'', and ``people''.
For each domain they were requested to answer $4$ existence test questions about the existence/nonexistence of some specific information in the schema and $4$ user experience questions.
Hence each domain collected $40$ to $52$ responses to existence test questions (shown in Table~\ref{tab:userstudy-conversion-size-all}), and $40$ to $52$ responses to user experience questions.

\vspace{-1mm}
\subsubsection{Existence Test Questions}

The existence test questions were designed to measure how helpful the various approaches are in assisting the participants to acquire a good understanding of the data.
An example existence test question is ``Based on this schema summary, I know the dataset provides the awards of a musician.''
The participants were requested to provide a Boolean yes/no answer.\vspace{-2mm}

{\flushleft \textbf{Time spent by participants:}}\hspace{2mm} We first verify if the approaches are convenient to use, in terms of how much time the participants must spend to answer the existence test questions.
For every existence test question that a participant worked on, we recorded the time spent by the participant on the question.
All participants worked on all the five domains in the same order.
As a participant gets gradually more familiar with the tasks, they tend to spend more time on the initial domains and less on the later domains.
This bias, due to budget and human resource constraints, makes it less meaningful to compare the time on existence tests across different domains.
A future study that allows every participant to work on only one domain can shed further light on how the complexity of a domain determines the time needed for its existence tests.

\begin{table}[t]
\centering
\scriptsize
\begin{tabular}{|l|l|l|l|l|l|}
\hline
 &  \textbf{books} & \textbf{film} & \textbf{music} & \textbf{TV} & \textbf{people} \\ \hline
\textbf{Concise} & n=52  & n=52 & n=52 & n=52 & n=52 \\
 &  c=0.730 & c=0.865 & c=0.903 & c=0.884 & c=0.788  \\ \hline
\textbf{Tight} & n=48 & n=48 & n=48 & n=48 & n=48\\
 & c=0.687 & c=0.854 & c=0.979 & c=0.875 & c=0.666  \\ \hline
\textbf{Diverse} & n=52 &  n=51 &  n=52 &  n=48 & n=48\\
 & c=0.846 & c=0.921 & c=0.730 & c=0.75 & c=0.875  \\ \hline
\textbf{Freebase} & n=44 & n=44 & n=44 & n=44 & n=44\\
 & c=0.818 & c=0.954  & c=0.931 & c=0.909 & c=0.681 \\ \hline
\textbf{Experts} & n=48 &  n=48 & n=48 & n=48 & n=48\\
 & c=0.604 & c=0.833 & c=0.895 & c=0.812 & c=0.687 \\ \hline
\textbf{YPS09} & n=52 & n=52 & n=52 & n=52 & n=52 \\
 & c=0.692  & c=0.884 & c=0.923 & c=0.692 & c=0.634 \\ \hline
\textbf{Graph} & n=40 & n=40 & n=40 & n=40 & n=40 \\
 & c=0.975 & c=0.875 & c=0.875 & c=0.9 & c=0.85\\ \hline
\end{tabular}
\vspace{-2mm}
\caption{\small Sample sizes and conversion rates for all approaches and domains. (For ``Diverse'' and ``film'', $51$ instead of $52$ responses were recorded. One response was lost, likely due to imperfect implementation of session management in the data collection website.)}\vspace{-3mm}
\label{tab:userstudy-conversion-size-all}
\end{table}

\begin{figure}[t]
\begin{center}
     \includegraphics[width=0.27\textwidth]{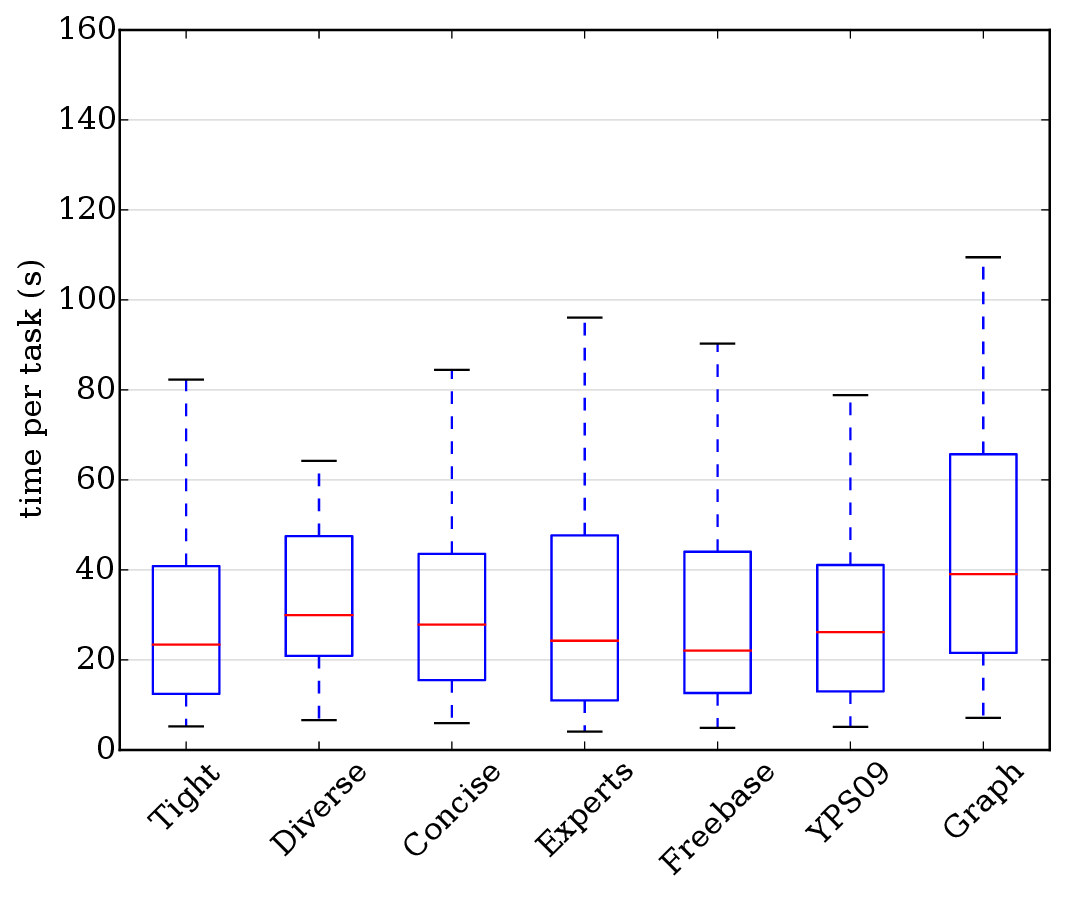}
\end{center} \vspace{-9mm}
\caption{\small Time taken on existence tests, domain=``music''.} \label{fig:time-bxplot-music}
\vspace{-1mm}
\end{figure}

\begin{table}[t]
\centering
\scriptsize
\begin{tabular}{|@{\hspace{0.35em}}l@{\hspace{0.35em}}|@{\hspace{0.35em}}l@{\hspace{0.35em}}|@{\hspace{0.35em}}l@{\hspace{0.35em}}|@{\hspace{0.35em}}l@{\hspace{0.35em}}|@{\hspace{0.35em}}l@{\hspace{0.35em}}|@{\hspace{0.35em}}l@{\hspace{0.35em}}|@{\hspace{0.35em}}l@{\hspace{0.35em}}|@{\hspace{0.35em}}l@{\hspace{0.35em}}|}
\hline
 \textbf{Domain} &  \textbf{1} & \textbf{2} & \textbf{3} & \textbf{4} & \textbf{5} & \textbf{6} & \textbf{7} \\ \hline
\textbf{books} & Graph & Freebase & Diverse &  Tight & Concise & YPS09 & Experts \\
\hline
\textbf{film} &  Tight & Freebase & Diverse & Concise & Experts & Graph & YPS09 \\
\hline
\textbf{music} &  Freebase & Tight & Experts & YPS09 & Concise & Diverse & Graph \\
\hline
\textbf{TV} & Tight  & YPS09  & Experts & Graph & Diverse & Concise & Freebase\\
\hline
\textbf{people} & Tight  & Freebase & Concise & Diverse  & Experts & YPS09 & Graph \\
\hline
\end{tabular}
\vspace{-2mm}
\caption{\small Systems sorted in ascending order by the median time spent on existence test questions.}\vspace{-3mm}
\label{tab:userstudy-sorted_time}
\end{table}

\begin{table}[t]
\centering
\scriptsize
{
\setlength{\tabcolsep}{0.35em}
\begin{tabular}{|l|l|l|l|l|l|l|}
\hline
\rowcolor {blue!50}
 & \textbf{Tight} & \textbf{Diverse} & \textbf{Freebase} & \textbf{Experts} & \textbf{YPS09} & \textbf{Graph} \\ \hline
\cellcolor {blue!15}\textbf{Concise} & z=\cellcolor {blue!50}1.59 &\cellcolor {blue!15}z=$-$2.28 & z=0.49 &z=$-$0.13 &z=0.36 &z=$-$0.43 \\
\cellcolor {blue!15}  &p=\cellcolor {blue!50}0.0559 &\cellcolor {blue!15}p=0.0113 &p=0.3121 &p=0.4483 &p=0.3594 &p=0.3336 \\
 \hline
\cellcolor {blue!15} \textbf{Tight} & &\cellcolor {blue!15}z=$-$3.48 & z=$-$1.12 &\cellcolor {blue!15}z=$-$1.69 &\cellcolor {blue!15}z=$-$1.282 &\cellcolor {blue!15}z=$-$1.93 \\
\cellcolor {blue!15} &   &\cellcolor {blue!15}p=0.0003 &p=0.1314  &\cellcolor {blue!15}p=0.0455 &\cellcolor {blue!15}p=0.0999 &\cellcolor {blue!15}p=0.0268 \\
\hline
\cellcolor {blue!15} \textbf{Diverse} &  & & \cellcolor {blue!50}z=2.57 &\cellcolor {blue!50}z=2.10 &\cellcolor {blue!50}z=2.60 &\cellcolor {blue!50}z=1.70 \\
\cellcolor {blue!15} &  &  &\cellcolor {blue!50}p=0.0051 &\cellcolor {blue!50}p=0.0179 &\cellcolor {blue!50}p=0.0047 &\cellcolor {blue!50}p=0.0446 \\
\hline
\cellcolor {blue!15} \textbf{Freebase} & & & &z=$-$0.61 &z=$-$0.15 &z=$-$0.87 \\
\cellcolor {blue!15} & & & &p=0.2709 &p=0.4404 &p=0.1922 \\
\hline
\cellcolor {blue!15} \textbf{Experts} &  & &  & &z=0.49 &z=$-$0.29\\
\cellcolor {blue!15} & & & & &p=0.3121 &p=0.3859\\
\hline
\cellcolor {blue!15} \textbf{YPS09} &  && & & &z=$-$0.77 \\
\cellcolor {blue!15} & && & &&p=0.2206\\
\hline
\end{tabular}
}
\vspace{-2mm}
\caption{\small Pairwise comparisons of seven approaches' conversion rates, domain=``music''.}
\label{tab:userstudy-z-p-music}
\end{table}

The time per question for domain ``music'' is displayed in the boxplots in Fig.~\ref{fig:time-bxplot-music}, and the results for other domains are included in the Appendix (Figs.~\ref{fig:time-bxplot-book} to~\ref{fig:time-bxplot-people}).
Table~\ref{tab:userstudy-sorted_time} provides a summary of the results.
For each domain, it sorts all seven approaches in ascending order by the median time spent by participants on the existence test questions.
Tight preview appears to be the most convenient approach, as its participants needed the least amount of time in three out of five domains and the second least in a fourth domain.
The Freebase gold standard also did well, as expected.
Surprisingly the previews produced by experts did not fare well.
This may indicate the challenges in generating truly useful previews by hands, even though the experts spent a lot of time.
``Diverse'' and ``Concise'' are ranked in the middle.
In general ``YPS09'' and ``Graph'' are the least convenient approaches.
For ``YPS09'', the table for each entity type includes all relationships incident on the entity type, as explained in Sec.~\ref{sec:yps09}.
Since~\cite{DBLP:journals/pvldb/YangPS09} only clusters the tables and does not discern the importance of different attributes for each table, the tables are wide.
Therefore they are less convenient in existence tests.
For ``Graph'', its inconvenience may not be difficult to understand, given the complexity of a schema graph.\vspace{-2mm}

{\flushleft \textbf{Accuracy of participants:}}\hspace{2mm}
We measured the effectiveness of the seven approaches by \emph{conversion rate}, which is the percentage of existence test questions correctly answered by the participants.
The conversion rates are shown in Table~\ref{tab:userstudy-conversion-size-all}.
Based on their values, we compare the seven approaches in a pairwise fashion.
Table~\ref{tab:userstudy-z-p-music} reports the results for the ``music'' domain.
The results for other domains can be found in Tables~\ref{tab:userstudy-z-p-books} to~\ref{tab:userstudy-z-p-people} in Appendix.
In the tables, each cell shows the hypothesis testing outcome when we compare the two approaches indicated by the corresponding column label and row label.
If a cell is in light blue, users of the approach corresponding to the cell's row label are more accurate in existence tests than users of the approach corresponding to the column label, and the outcome is statistically significant.
If a cell is in dark blue, it is the opposite.
If a cell is not colored, we cannot make a statistically significant conclusion regarding which of the two approaches leads to more accurate users.
Below we explain the hypothesis testing in more detail.

\begin{table*}[t]
\centering
\scriptsize
\begin{tabular}{c|l|l|l|l|}
\hline
\multicolumn{1}{|l|}{\parbox{13mm}{\textbf{Likert Scale Score}}} &   \parbox{25mm}{\textbf{Q1:} How easy was it to read the schema summary of this domain?}   &   \parbox{34mm}{\textbf{Q2:} How much understanding of the data in this domain can you gain from the schema summary?}  &   \parbox{34mm}{\textbf{Q3:} How helpful was the schema summary in assisting you to understand the data of this domain?}  &  \parbox{42mm}{\textbf{Q4:} Is the schema summary missing important information about data in this domain?}  \\ \hline
\multicolumn{1}{|l|}{1} & Very hard     & Very little  &  Not helpful at all & It provides very little important information.   \\
\multicolumn{1}{|l|}{2} &   Hard    & A Little  &Did not help much &  It provides some important information.   \\
\multicolumn{1}{|l|}{3} &   Neutral   & Neutral & Neutral & Neutral   \\
\multicolumn{1}{|l|}{4} &   Easy     & Some  &  Somewhat helpful &  It provides most of the important information.   \\
\multicolumn{1}{|l|}{5} &   Very easy  & Very much & Very helpful & It provides all important information.   \\ \hline
\end{tabular}
\vspace{-3mm}
\caption{\small User experience questionnaire.}
\vspace{-4mm}
\label{tab:survey}
\end{table*}

Each cell shows a $z$-score and a $p$-value, which are the outcomes of a two-proportion one-tailed $z$-test with significance level $\alpha$$=$$0.1$.
Such a hypothesis testing is proper, since our samples (responses from participants using different approaches) are independent and the sample sizes are large enough.
Consider a cell at the intersection of column A and row B.
The hypothesis testing for the difference between the two proportions for A and B is as follows.
We assume that answering the existence test questions follows a Bernoulli trial with the probabilities of success $p_A$ and $p_B$ for approaches A and B, respectively.
The observed conversion rates of A and B, $c_A$ and $c_B$, are in Table~\ref{tab:userstudy-conversion-size-all}.
For $c_A$$>$$c_B$ (resp., $c_A$$<$$c_B$), the null hypothesis is $H_0$: $p_A$$\leq$$p_B$ (resp., $p_A$$\geq$$p_B$) and the alternative hypothesis is $H_a$: $p_A$$>$$p_B$ (resp., $p_A$$<$$p_B$).
According to the sample sizes ($n_A$ and $n_B$) and observed conversion rates ($c_A$ and $c_B$) in Table~\ref{tab:userstudy-conversion-size-all}, we calculate the $z$-score.
For calculating the $p$-value, if the $z$-score is positive (i.e., $c_A$$>$$c_B$), we use a right-tailed $z$-test; otherwise we use a left-tailed $z$-test.
Suppose the $p$-value is less than $\alpha$.
Then $H_0$ will be rejected and the data significantly supports the claim that users of A (resp., B) have a higher chance of answering existence tests correctly, if $c_A$$>$$c_B$ (resp., $c_B$$>$$c_A$).

The hypothesis testing outcomes for different domains exhibit certain degree of diversity.
In domain ``music'' (Table~\ref{tab:userstudy-z-p-music}), ``Tight'' outperformed all but ``Freebase''.
In comparison with ``Freebase'', the conversion rate of ``Tight'' is actually higher, although we cannot reject the null hypothesis.
On the other hand, ``Diverse'' performed poorly in this domain, as it is statistically significantly worse than all other approaches.
In domain ``books'' (Table~\ref{tab:userstudy-z-p-books}), ``Graph'' had the best performance, and ``Diverse'' did well too.
In this domain ``Tight'' and ``Experts'' did poorly.
In domain ``film'', ``Freebase'' did well (Table~\ref{tab:userstudy-z-p-film}).
In domain ``TV'', ``YPS09'' had the worst performance and no approach positively stood out (Table~\ref{tab:userstudy-z-p-tv}).
In domain ``people'' (Table~\ref{tab:userstudy-z-p-people}), both ``Graph'' and ``Diverse'' performed very well.
Across all domains, it is quite surprising that ``Experts'' was never statistically significantly better than any other approach, except for ``Diverse'' in domain ``music''.

\begin{table}[t]
\centering
\scriptsize
\begin{tabular}{|@{\hspace{0.35em}}l@{\hspace{0.35em}}|@{\hspace{0.35em}}l@{\hspace{0.35em}}|@{\hspace{0.35em}}l@{\hspace{0.35em}}|@{\hspace{0.35em}}l@{\hspace{0.35em}}|@{\hspace{0.35em}}l@{\hspace{0.35em}}|@{\hspace{0.35em}}l@{\hspace{0.35em}}|@{\hspace{0.35em}}l@{\hspace{0.35em}}|@{\hspace{0.35em}}l@{\hspace{0.35em}}|}
\hline
 \textbf{Question} &  \textbf{1} & \textbf{2} & \textbf{3} & \textbf{4} & \textbf{5} & \textbf{6} & \textbf{7} \\ \hline
\textbf{Q1} & Freebase & Diverse & Graph & Experts & YPS09 & Concise & Tight \\
\hline
\textbf{Q2} & Graph	& Freebase & YPS09 & Diverse & Concise & Tight & Experts \\
\hline
\textbf{Q3} & Graph	& Freebase & YPS09 & Diverse & Experts & Concise & Tight \\
\hline
\textbf{Q4} & YPS09	& Concise & Experts & Graph	& Tight & Freebase & Diverse \\
\hline
\end{tabular}
\vspace{-3mm}
\caption{\small Systems sorted in descending order by average user experience scores across five domains.}\vspace{-1mm}
\label{tab:userstudy-sorted_score}
\end{table}

\vspace{-2mm}
\subsubsection{User Experience Questions}
In each domain, we asked every participant four user experience questions, after the four existence test questions.
The four questions Q1--Q4 are listed in Table~\ref{tab:survey}.
Each question comes with five options, specifying the level of satisfaction a participant
may have regarding the particular aspect of the approach measured by the question.
We assign a score to every option, based on the Likert scale shown in Table~\ref{tab:survey}.
The least favourable experience with respect to each question is assigned a score of $1$,
and the most favourable experience is assigned a score of $5$.
For a certain approach, the overall user experience score for each question is measured by averaging the scores obtained for that question from all the participants using that approach.

Results for individual domains can be found in the Appendix  (Tables~\ref{tab:experience-score-book} to \ref{tab:experience-score-people}).
The results for different domains are diverse, likely due to their different sizes and complexities.
Hence, we summarize the results in Table~\ref{tab:userstudy-sorted_score}.
For each user experience question, Table~\ref{tab:userstudy-sorted_score} sorts all seven approaches in descending order by their average user experience scores across all five domains.
Overall, the results suggest a mismatch between the participants' perception and their efficacy in answering existence test questions.
The only exception appears to be ``Freebase'', of which the participants' perception largely agrees with their performance in using the approach.
This may not be surprising, given that ``Freebase'' is the gold standard.
Regarding Q1, while Table~\ref{tab:userstudy-sorted_time} indicates that ``Tight'' is the most convenient approach, the participants' perception suggests the opposite.
Regarding Q2 and Q3, although the hypothesis testing results discussed earlier favor ``Tight'' in many situations, it once again did not fare well in leaving a satisfactory impression on the participants.
The participants believed they acquired more understanding of the data when they used ``Graph'' and ``YPS09'', although the hypothesis testing results suggest that they typically answered the existence test questions more accurately when they use approaches such as ``Tight''.
Regarding Q4, it is interesting that the participants favored ``YPS09'' the most, although they answered the questions less accurately using ``YPS09'' than using approaches such as ``Tight''.
A logical explanation to these mismatches might be that the more complex presentation used in ``Graph'' and ``YPS09'' triggered the participants to believe that they had better understanding of the data and they had seen more complete information.
A similar observation was made regarding ``Tight'' and ``Diverse''---``Tight'' clearly helped participants to do existence tests more accurately and quickly, but the participants had better impression of ``Diverse''.
More thorough and robust explanation of these observations is the goal of future investigation, which likely will need to involve larger-scale user study and in-person interviews.

\vspace{-2mm}

\section{Related Work}
\label{sec:relatedworks}
There have been several studies on schema summarization for relational databases~\cite{DBLP:journals/pvldb/YangPS09, DBLP:journals/pvldb/YangPS11a, DBLP:conf/vldb/YuJ06a}, XML~\cite{DBLP:conf/vldb/YuJ06a} and general graph data~\cite{DBLP:conf/sigmod/TianHP08, DBLP:conf/icde/ZhangTP10}.
\cite{DBLP:conf/vldb/YuJ06a} produces schema summarization for relational databases and XML data.
The notion of summary in~\cite{DBLP:journals/pvldb/YangPS09,DBLP:journals/pvldb/YangPS11a} refers to clustering the tables in a database by their semantic roles and similarities as well as identifying direct join relationships and indirect join paths between the tables.
The graph summarization in~\cite{DBLP:conf/sigmod/TianHP08,DBLP:conf/icde/ZhangTP10} groups graph nodes based on their attribute similarity and allows users to browse the summary from different grouping granularities.
As explained in Sec.~\ref{sec:introduction}, these methods are inapplicable or ineffective for producing preview tables from entity graphs, due to differences in input/output data models and goals.

There are many works on graph clustering~\cite{Schaeffer:2007:SGC:2296006.2296057}.  They are not effective for generating preview tables, since clustering focuses on partitioning but does not present a concise structure.  On the contrary, a preview only selects a small number of key attributes (vertices) and non-key attributes (edges) from a schema graph.

\cite{qunits} proposed the concept of queried units (``qunits'') for representing desired query results on a database.  For automatic derivation of qunits, \cite{qunits} discussed several ideas.  One idea is to utilize the concept of queriability~\cite{formquery} which measures the importance of a schema entity by its schema connectedness and its data cardinality.  The measure is thus similar to our key attribute scoring measures (Sec.~\ref{sec:key-score}).  ObjectRank~\cite{objectrank} applies authority-based ranking to keyword search in databases.  Part of its ranking formula is extended from PageRank.  The table importance measure in~\cite{DBLP:journals/pvldb/YangPS09} and our random-walk based scoring measure (Sec.~\ref{sec:key-score}) bear similar ideas.

\cite{snippetxml} studied how to generate query result snippets in XML search. Similar to~\cite{DBLP:conf/vldb/YuJ06a}, they focus on semi-structured data.  Differently, they produce snippets of query results while~\cite{DBLP:conf/vldb/YuJ06a} summarizes schema.  In \cite{snippetxml}, the problem of generating snippets is formulated as maximizing information under an upper bound on snippet size.  At high level, this is similar to our problem of finding optimal previews under size constraint, although its detailed problem formulation, solution, and data model are different.

\vspace{-2mm}

\section{Conclusion}
\label{sec:conclusion}
\vspace{-1mm}
This paper studies how to generate preview tables for entity graphs.  The problem is challenging due to the scale and complexity of such graphs.  We proposed effective scoring measures for preview tables.  We proved that the optimal preview discovery problem under distance constraint is \NP-hard.  We designed efficient algorithms for discovering optimal previews. The experiments and user study verified the effectiveness of our methods.

There can be several future directions worth pursuing. (1) Guidelines and automatic techniques for choosing between tight and diverse previews. (2) Selecting representative entity tuples for preview tables.
(3) Incorporating numeric attributes into preview tables.
(4) Suggesting values of various parameters, including $N$, $K$ and distance constraints for tight and diverse previews.

\vspace{-2mm}
{\small {\flushleft \textbf{Acknowledgments}\hspace{1mm}} The authors have been partially supported by NSF grants 1018865, 1408928 and NSF-China grant 61370019. Any opinions, findings, and conclusions in this publication are those of the authors and do not necessarily reflect the views of the funding agencies. We also thank Rudresh Ajgaonkar and Aaditya Kulkarni for contributions in user study.}

\clearpage
\bibliographystyle{abbrv}

\begin{thebibliography}{10}

\bibitem{apriori}
R.~Agarwal and R.~Srikant.
\newblock Fast algorithms for mining association rules.
\newblock In {\em VLDB}, pages 487--499, 1994.

\bibitem{AuerBK+07}
S.~Auer, C.~Bizer, G.~Kobilarov, J.~Lehmann, R.~Cyganiak, , and Z.~Ives.
\newblock {DB}pedia: A nucleus for a {Web} of open data.
\newblock In {\em ISWC}, pages 722--735, 2007.

\bibitem{objectrank}
A.~Balmin, V.~Hristidis, and Y.~Papakonstantinou.
\newblock Objectrank: Authority-based keyword search in databases.
\newblock In {\em VLDB}, pages 564--575, 2004.

\bibitem{Bollacker+08freebase}
K.~Bollacker, C.~Evans, P.~Paritosh, T.~Sturge, and J.~Taylor.
\newblock Freebase: a collaboratively created graph database for structuring
  human knowledge.
\newblock In {\em SIGMOD}, pages 1247--1250, 2008.

\bibitem{Brin:1998:ALH:297805.297827}
S.~Brin and L.~Page.
\newblock The anatomy of a large-scale hypertextual web search engine.
\newblock In {\em WWW}, pages 107--117, 1998.

\bibitem{Bron-Kerbosch}
C.~Bron and J.~Kerbosch.
\newblock Algorithm 457: finding all cliques of an undirected graph.
\newblock {\em CACM}, 16(9):575--577, Sept. 1973.

\bibitem{pcc_cohen}
J.~Cohen.
\newblock {\em {Statistical Power Analysis for the Behavioral Sciences}}.
\newblock Academic Press, 1988.

\bibitem{knowledgevault}
X.~Dong, E.~Gabrilovich, G.~Heitz, W.~Horn, N.~Lao, K.~Murphy, T.~Strohmann,
  S.~Sun, and W.~Zhang.
\newblock Knowledge vault: A web-scale approach to probabilistic knowledge
  fusion.
\newblock In {\em KDD}, pages 601--610, 2014.

\bibitem{snippetxml}
Y.~Huang, Z.~Liu, and Y.~Chen.
\newblock Query biased snippet generation in xml search.
\newblock In {\em SIGMOD}, pages 315--326, 2008.

\bibitem{formquery}
M.~Jayapandian and H.~V. Jagadish.
\newblock Automated creation of a forms-based database query interface.
\newblock {\em PVLDB}, 1(1):695--709, Aug. 2008.

\bibitem{journals/bioinformatics/KoseWLF01}
F.~Kose, W.~Weckwerth, T.~Linke, and O.~Fiehn.
\newblock Visualizing plant metabolomic correlation networks using
  clique-metabolite matrices.
\newblock {\em Bioinformatics}, 17(12):1198--1208, Dec. 2001.

\bibitem{learn-to-rank}
T.-Y. Liu.
\newblock Learning to rank for information retrieval.
\newblock {\em Found. Trends Inf. Retr.}, 3(3):225--331, Mar. 2009.

\bibitem{Manning08}
C.~D. Manning, P.~Raghavan, and H.~Schtze.
\newblock {\em Introduction to Information Retrieval}.
\newblock Cambridge University Press, 2008.

\bibitem{qunits}
A.~Nandi and H.~V. Jagadish.
\newblock Qunits: queried units in database search.
\newblock In {\em CIDR}, 2009.

\bibitem{Schaeffer:2007:SGC:2296006.2296057}
S.~E. Schaeffer.
\newblock Survey: Graph clustering.
\newblock {\em Comput. Sci. Rev.}, 1(1):27--64, Aug. 2007.

\bibitem{SuchanekKW07}
F.~M. Suchanek, G.~Kasneci, and G.~Weikum.
\newblock {YAGO}: a core of semantic knowledge unifying {WordNet} and
  {Wikipedia}.
\newblock In {\em WWW}, pages 697--706, 2007.

\bibitem{DBLP:conf/sigmod/TianHP08}
Y.~Tian, R.~A. Hankins, and J.~M. Patel.
\newblock Efficient aggregation for graph summarization.
\newblock In {\em SIGMOD}, pages 567--580, 2008.

\bibitem{probase}
W.~Wu, H.~Li, H.~Wang, and K.~Q. Zhu.
\newblock Probase: a probabilistic taxonomy for text understanding.
\newblock In {\em SIGMOD}, pages 481--492, 2012.

\bibitem{DBLP:journals/pvldb/YangPS09}
X.~Yang, C.~M. Procopiuc, and D.~Srivastava.
\newblock Summarizing relational databases.
\newblock {\em PVLDB}, 2(1):634--645, 2009.

\bibitem{DBLP:journals/pvldb/YangPS11a}
X.~Yang, C.~M. Procopiuc, and D.~Srivastava.
\newblock Summary graphs for relational database schemas.
\newblock {\em PVLDB}, 4(11):899--910, 2011.

\bibitem{DBLP:conf/vldb/YuJ06a}
C.~Yu and H.~V. Jagadish.
\newblock Schema summarization.
\newblock In {\em VLDB}, pages 319--330, 2006.

\bibitem{DBLP:conf/icde/ZhangTP10}
N.~Zhang, Y.~Tian, and J.~M. Patel.
\newblock Discovery-driven graph summarization.
\newblock In {\em ICDE}, pages 880--891, 2010.

\end{thebibliography}

\newpage

\appendix

\section{Schemata of the Tables in the Freebase Gold Standard}
The schema of the tables in the ``Freebase'' gold standard can be found in Table~\ref{tab:appendix-gold-standard}.

\begin{table}[h]
\centering
\scriptsize
    \begin{tabular}{|l|l|}
    \hline
    \textbf{Key attributes}  & \textbf{Non-key attributes}   \\ \hline \hline

    \multicolumn{2}{ |c| }{ Domain=``books'', $k$=$6$, $n$=$15$} \\ \hline
    \etype{Book} & \edge{Characters},  \edge{Genre},  \edge{Editions} \\
    \etype{Book Edition}   & \edge{Publication Date}, \edge{Publisher}, \edge{Credited To } \\
    \etype{Short Story}     & \edge{Genre}, \edge{Characters}  \\
    \etype{Poem}    & \edge{Characters}, \edge{Meter}, \edge{Verse Form}  \\
    \etype{Short Non-fiction}     & \edge{Mode Of Writing}, \edge{Verse Form}  \\
    \etype{Author}     & \edge{Series Written (Or Contributed To)},  \edge{Works Edited}, \\
       &  \edge{Works Written}  \\
    \hline \hline
    \multicolumn{2}{ |c| }{ Domain=``film'', $k$=$6$, $n$=$9$}  \\ \hline
    \etype{Film}  & \edge{Directed By}, \edge{Tagline}, \edge{Initial Release Date}  \\
    \etype{Film Actor}      & \edge{Film performances}    \\
    \etype{Film genre}      & \edge{Films of this genre} \\
    \etype{Film Director}   & \edge{Films directed}   \\
    \etype{Film producer} & \edge{Films Executive Produced}, \edge{Films Produced}      \\
     \etype{Film writer}   & \edge{Film Writing Credits}   \\
   \hline \hline
   \multicolumn{2}{ |c| }{ Domain=``music'', $k$=$6$, $n$=$18$} \\ \hline
    \etype{Composition} & \edge{Includes}, \edge{Lyricist}, \edge{Composer}  \\
    \etype{Concert}   & \edge{Venue}, \edge{Start Date}, \edge{Concert Tour}  \\
    \etype{Music video}     & \edge{Song}, \edge{Initial release date}, \edge{Artist}  \\
    \etype{Musical Album}   & \edge{Release Type}, \edge{Initial Release Date}, \edge{Artist}  \\
    \etype{Musical Artist}     & \edge{Albums}, \edge{Place Musical Career Began},  \\
    & \edge{Musical Genres}  \\
     \etype{Musical Recording}     & \edge{Length}, \edge{Featured artists}, \edge{Recorded by}  \\
     \hline     \hline
    \multicolumn{2}{ |c| }{ Domain=``TV'', $k$=$6$, $n$=$9$} \\ \hline
    \etype{TV Program} & \edge{Program Creator}, \edge{Air Date Of First Episode},  \\
    &   \edge{Air Date Of Final Episode}  \\
    \etype{TV Actor}   & \edge{Starring TV Roles}  \\
    \etype{TV Character}     & \edge{Programs In Which This Was A Regular Character} \\
    \etype{TV Writer}     & \edge{TV Programs (Recurring Writer)} \\
    \etype{TV Producer}     & \edge{TV Programs Produced} \\
    \etype{TV Director}     & \edge{TV Episodes Directed}, \edge{TV Segments Directed}  \\
    \hline     \hline
 \multicolumn{2}{ |c| }{ Domain=``people'', $k$=$6$, $n$=$16$} \\ \hline
   \etype{Person} & \edge{Profession}, \edge{Country Of Nationality}, \edge{Date Of Birth}  \\
    \etype{Deceased Person}   & \edge{Cause Of Death}, \edge{Place Of Death}, \edge{Date Of Death}  \\
    \etype{Cause Of Death}     & \edge{People Who Died This Way}, \\
    & \edge{Includes Causes Of Death}, \edge{Parent Cause Of Death}  \\
    \etype{Ethnicity}     & \edge{Geographic Distribution}, \edge{Includes Group(S)},  \\
    & \edge{Included In Group(S)}  \\
    \etype{Profession}     & \edge{Specializations}, \edge{Specialization Of},  \\
    & \edge{People With This Profession}  \\
     \etype{Professional field}     & \edge{Professions In This Field} \\
\hline
    \end{tabular}
\caption{\small Gold standard (``Freebase''). For each domain, there are 6 key attributes and at most 3 non-key attributes for each key-attribute.}
\label{tab:appendix-gold-standard}
\end{table}

\begin{table}[h]
\centering
\scriptsize
    \begin{tabular}{|l|l|}
    \hline
    \textbf{Key attributes}  & \textbf{Non-key attributes (Target entity types)}  \\ \hline \hline

    \multicolumn{2}{ |c| }{ Domain=``film'', KS=Coverage, NKS=Coverage, $k$=$5$, $n$=$10$}  \\ \hline
    \etype{Film Character}  & \edge{Portrayed in films} (\etype{Film}, \etype{Film Actor})      \\
    \etype{Film Actor}      & \edge{Film performances} (\etype{Film}, \etype{Film Character})   \\
    \etype{Film}            & \edge{Performances} (\etype{Film actor}, \etype{Film Character}), \\
                    & \edge{Genres} (\etype{Film Genre}), \\
                    & \edge{Runtime} (\etype{Film Cut}), \\
                    & \edge{Country of origin} (\etype{Country}), \\
                    &  \edge{Directed by} (\etype{Film Director}), \\
                    & \edge{Languages} (\etype{Human Language}) \\
    \etype{Film Director}   & \edge{Films directed} (\etype{Film})  \\
    \etype{Film Crewmember} & \edge{Films crewed} (\etype{Film}, \etype{Film crew role}) \\ \hline \hline

    \multicolumn{2}{ |c| }{ Domain=``music'', KS=Random Walk, NKS=Coverage, $k$=$5$, $n$=$10$} \\ \hline
    \etype{Musical Recording} & \edge{Releases} (\etype{Musical Release}), \\
    & \edge{Tracks} (\etype{Release Track}), \\
    &   \edge{Recorded by} (\etype{Musical Artist}) \\
    \etype{Musical Release}   & \edge{Tracks} (\etype{Musical Recording}), \\
    & \edge{Track list} (\etype{Release Track})   \\
    \etype{Release Track}     & \edge{Release} (\etype{Musical Release}), \\
    & \edge{Recording} (\etype{Musical Recording})  \\
    \etype{Musical Artist}    & \edge{Tracks recorded} (\etype{Musical Recording})   \\
    \etype{Musical Album}     & \edge{Releases} (\etype{Musical Release}), \\
    &   \edge{Release type} (\etype{Musical Album Type})   \\ \hline     \hline

    \multicolumn{2}{ |c| }{ Domain=``TV'', KS=Random Walk, NKS=Entropy, $k$=$5$, $n$=$10$} \\ \hline

    \etype{TV Episode} & \edge{Previous episode} (\etype{TV Episode}),  \\
    & \edge{Next episode} (\etype{TV Episode}), \\
     &   \edge{Performances} (\etype{TV Actor}, \etype{TV Character}), \\
     &   \edge{Season} (\etype{TV Season}),\\
     &  \edge{Series} (\etype{TV Program}) , \\
     &  \edge{Personal appearances} \\
     & (\etype{Person}, \etype{Personal Appearance Role}) \\
	\etype{TV Program} & \edge{Regular acting performances} \\
	& (\etype{TV Actor}, \etype{TV Character}, \etype{TV Season}) \\
	\etype{TV Season} & \edge{Episodes} (\etype{TV Episode}) \\
	\etype{TV Actor} & \edge{TV episode performances}\\
	&  (\etype{TV Episode}, \etype{TV Character}) \\
	\etype{TV Director} & \edge{TV episodes directed} (\etype{TV Episode}) \\ \hline
    \end{tabular}
\caption{\small Sample optimal concise previews.}\vspace{2mm}
\label{tab:samples1}
\end{table}

\begin{table}[h]
\centering
\scriptsize
    \begin{tabular}{|l|l|}
    \hline
    \textbf{Key attributes} & \textbf{Non-key attributes (Target entity types)}  \\ \hline \hline

    \multicolumn{2}{ |c| }{ Domain=``film'', KS=Coverage, NKS=Coverage, $k$=$5$, $n$=$10$, $d$=$2$} \\ \hline
    \etype{Film}          & \edge{Performances} (\etype{Film Character}, \etype{Film Actor}), \\
    & \edge{Genres} (\etype{Film Genre}), \\
    & \edge{Runtime} (\etype{Film Cut}),\\
    & \edge{Country of origin} (\etype{Country}),\\
    &  \edge{Directed by} (\etype{Film Director}),  \\
    & \edge{Languages} (\etype{Human Language}) \\
    \etype{Film Director} & \edge{Films directed} (\etype{Film})   \\
    \etype{Film Producer} & \edge{Films produced} (\etype{Film})     \\
    \etype{Film Writer}   & \edge{Film writing credits} (\etype{Film})  \\
    \etype{Film Editor}   & \edge{Films edited} (\etype{Film})    \\ \hline     \hline

    \multicolumn{2}{ |c| }{ Domain=``film'', KS=Coverage, NKS=Coverage, $k$=$5$, $n$=$10$, $d$=$4$}  \\ \hline
    \etype{Film Character}                    & \edge{Portrayed in films} (\etype{Film}, \etype{Film Actor}),  \\
    &  \edge{Portrayed in films (dubbed)} (\etype{Film}, \etype{Film Actor})                                              \\
    \etype{Film Crewmember}                    & \edge{Films crewed} (\etype{Film}, \etype{Film Crew Role})     \\
    \etype{Person or Entity}  & \edge{Films appeared in} (\etype{Film}, \etype{Type of Appearance}) \\
    \etype{appearing in Film} & \\
    \etype{Film Festival}                      & \edge{Individual festivals} (\etype{Film Festival Event}),  \\
    &  \edge{Location} (\etype{Location}),  \\
    & \edge{Focus} (\etype{Film Festival Focus}), \\
    &   \edge{Sponsoring organization} (\etype{Sponser}) \\
    \etype{Film Company}                       & \edge{Films} (\etype{Film})   \\	 \hline
    \end{tabular}
\caption{\small Sample optimal tight (upper) and diverse previews (lower).}\vspace{2mm}
\label{tab:samples2}
\end{table}

\vspace{-2mm}
\section{Sample Optimal Previews}
\label{sec:exp-sample}

To demonstrate the combined effectiveness of both scoring measures and preview discovery algorithms, Table~\ref{tab:samples1} presents the optimal concise previews in 3 selected domains by 3 different combinations of key attribute scoring (KS) and non-key attribute scoring (NKS) measures.  The size constraint is set as $k$=$5$ and $n$=$10$.  All result previews show that the selected key and non-key attributes have covered important entity types and their important relationship types.
Further, Table~\ref{tab:samples2} shows the optimal tight ($d$=$2$) and diverse ($d$=$4$) previews in ``film'' domain by one particular choice of key and non-key attribute scoring measures.
We see that, in the tight preview result, the chosen key attributes are all highly related to one entity type \etype{Film}.  In the diverse preview result, the chosen key attributes are far less related to each other.  Both verify the effectiveness of the concepts of tight/diverse previews.

Note that in the generated previews, certain non-key attributes represent relationship types involving more than two entity types.  An example in Table~\ref{tab:samples1} is \edge{Portrayed in films}, which is a non-key attribute of entity type \etype{Film Character}.  Different from other non-key attribute such as \edge{Films directed}, it represents a 3-way relationship among \etype{Film Character}, \etype{Film} and \etype{Film Actor}. For instance, \entity{Agent J} is a \etype{Film Character} played by \etype{Film Actor} \entity{Will Smith} in \etype{Film} \entity{Men in Black}.
To present the values of such a \textit{multi-way} non-key attribute in a preview table, we employ a simple approach of presenting values for all participating entity types in this relationship.  It is arguable that this approach widens the preview table, which to some extent violates a given size constraint.  An alternative solution is to use separate preview tables for all multi-way relationships. These pose interesting directions for our future work.

\section{Additional User Study Results}

\begin{figure}[h]
\begin{center}
     \includegraphics[width=0.3\textwidth]{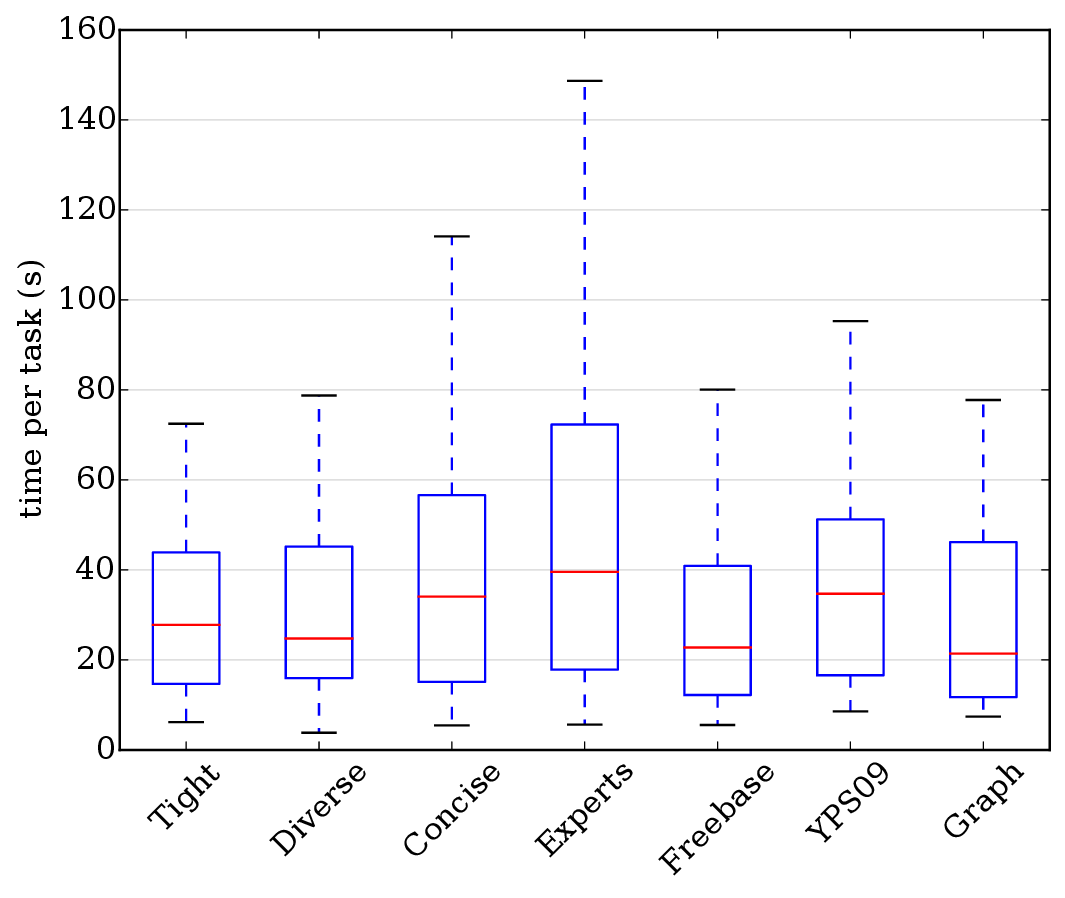}
\end{center} \vspace{-7mm}
\caption{\small Time taken on existence tests, domain=``books''.} \label{fig:time-bxplot-book}\vspace{-2mm}
\end{figure}
\begin{figure}[h]
\begin{center}
     \includegraphics[width=0.3\textwidth]{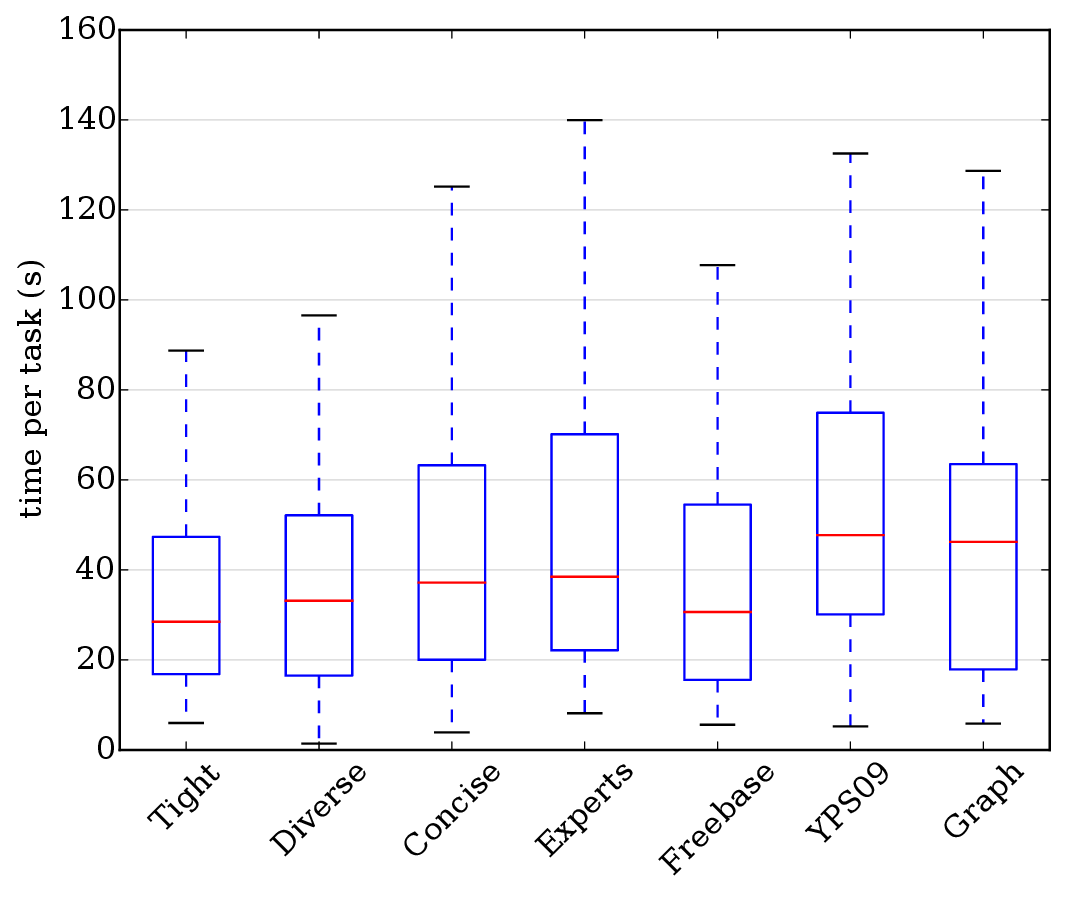}
\end{center} \vspace{-7mm}
\caption{\small Time taken on existence tests, domain=``film''.} \label{fig:time-bxplot-film}\vspace{-2mm}
\end{figure}
\begin{figure}[h]
\begin{center}
     \includegraphics[width=0.3\textwidth]{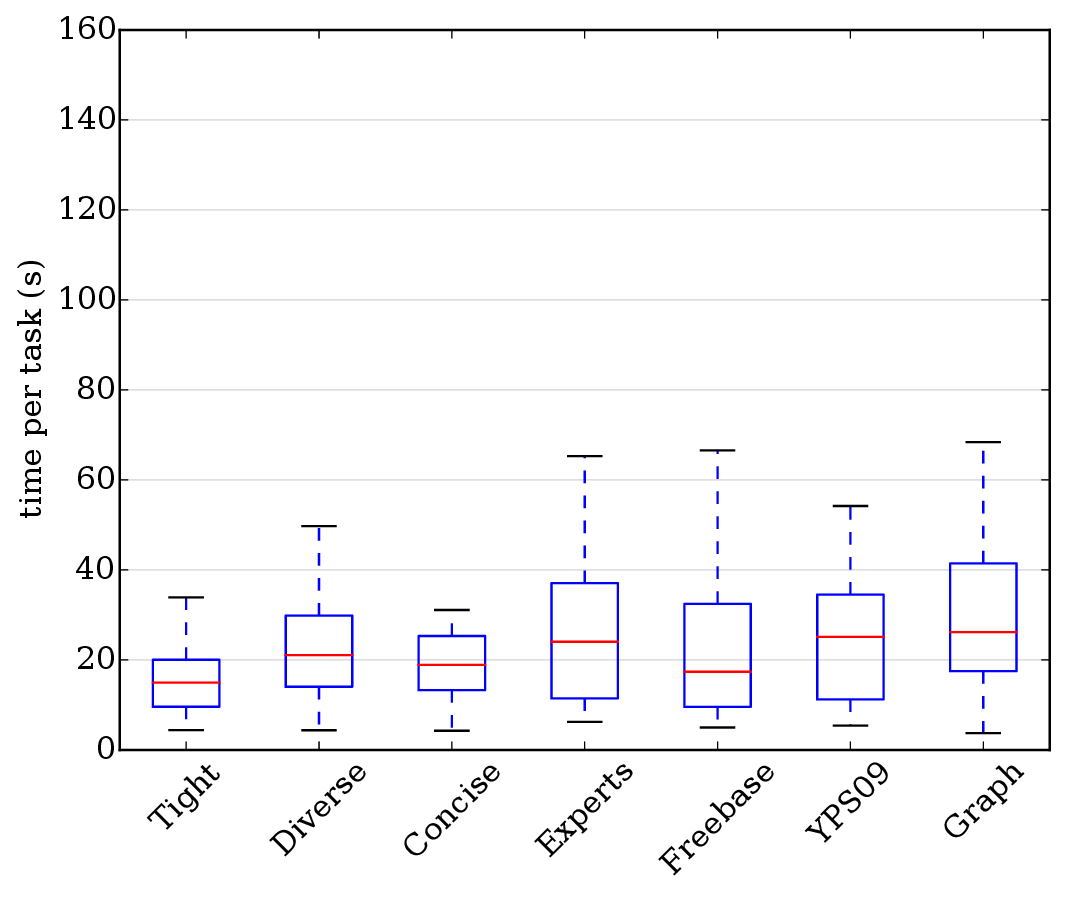}
\end{center} \vspace{-7mm}
\caption{\small Time taken on existence tests, domain=``TV''.} \label{fig:time-bxplot-tv}\vspace{-2mm}
\end{figure}
\begin{figure}[h]
\begin{center}
     \includegraphics[width=0.3\textwidth]{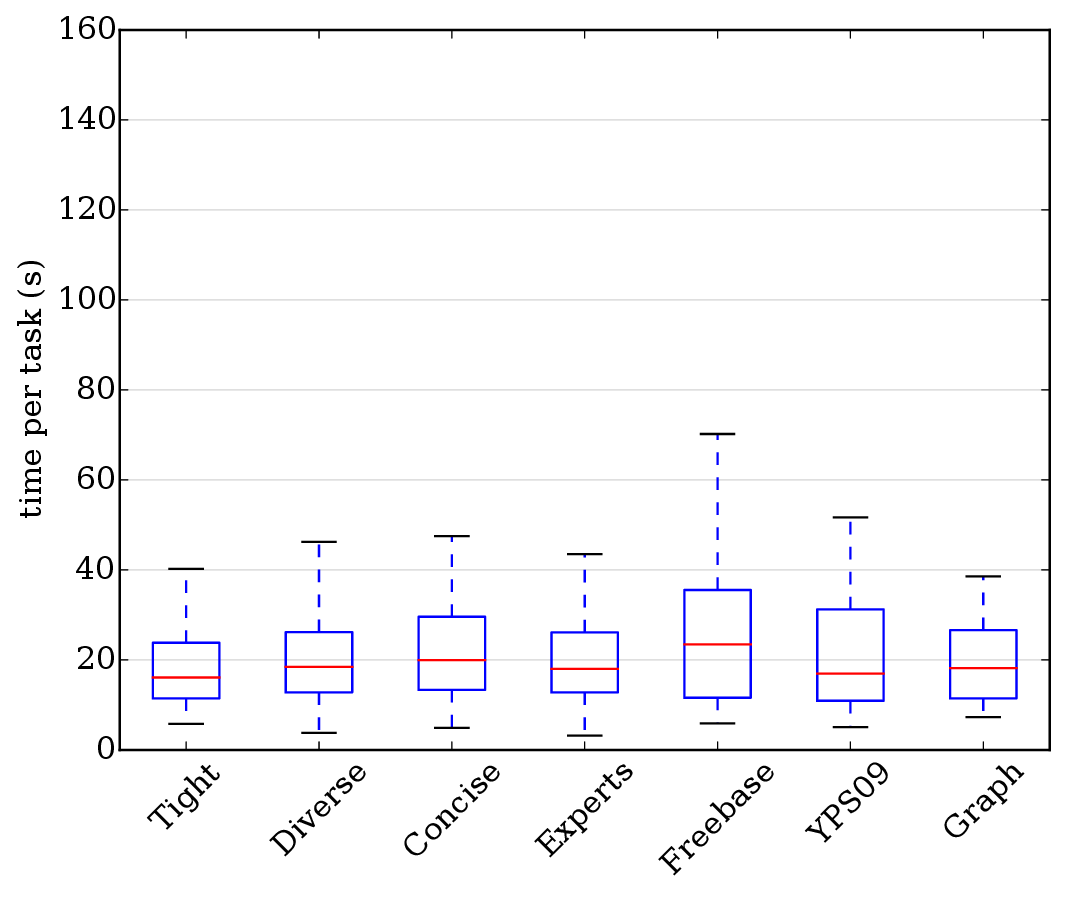}
\end{center} \vspace{-7mm}
\caption{\small Time taken on existence tests, domain=``people''.} \label{fig:time-bxplot-people}\vspace{-2mm}
\end{figure}

\begin{table}[h]
\centering
\scriptsize
{
\setlength{\tabcolsep}{0.35em}
\begin{tabular}{|l|l|l|l|l|l|l|}
\hline
\rowcolor {blue!50}
 & \textbf{Tight} & \textbf{Diverse} & \textbf{Freebase} & \textbf{Experts} & \textbf{YPS09} & \textbf{Graph} \\ \hline
\cellcolor {blue!15}\textbf{Concise} & z=$-$0.47 &\cellcolor {blue!50}z=1.45 & z=1.02 &\cellcolor {blue!15}z=$-$1.34 &z=$-$0.43 &\cellcolor {blue!50}z=3.15 \\
\cellcolor {blue!15} & p=0.3192 &\cellcolor {blue!50}p=0.0735 &p=0.1539  &\cellcolor {blue!15}p=0.0901 &p=0.3336 &\cellcolor {blue!50}p=0.0008 \\
 \hline
\cellcolor {blue!15} \textbf{Tight} & &\cellcolor {blue!50}z=1.89 &\cellcolor {blue!50} z=1.45 &z=$-$0.85 &z=0.05 &\cellcolor {blue!50}z=3.49 \\
\cellcolor {blue!15} & &\cellcolor {blue!50}p=0.0294 &\cellcolor {blue!50}p=0.0735 &p=0.1977 &p=0.4801 &\cellcolor {blue!50}p=0.0002 \\
\hline
\cellcolor {blue!15} \textbf{Diverse} &  & & z=$-$0.37 &\cellcolor {blue!15}z=$-$2.72 &\cellcolor {blue!15}z=$-$1.86 &\cellcolor {blue!50}z=2.06 \\
\cellcolor {blue!15} & & & p=0.3557 &\cellcolor {blue!15}p=0.0033  &\cellcolor {blue!15}p=0.0314 &\cellcolor {blue!50}p=0.0197 \\
\hline
\cellcolor {blue!15} \textbf{Freebase} &  & & &\cellcolor {blue!15}z=$-$2.25 &\cellcolor {blue!15}z=$-$1.42 &\cellcolor {blue!50}z=2.32 \\
\cellcolor {blue!15} & & & & \cellcolor {blue!15}p=0.0122  &\cellcolor {blue!15}p=0.0778  &\cellcolor {blue!50}p=0.0102 \\
\hline
\cellcolor {blue!15} \textbf{Experts} &  & &  & &z=0.92 &\cellcolor {blue!50}z=4.13 \\
\cellcolor {blue!15} & & & & &p=0.1788 &\cellcolor {blue!50}p=0.0000 \\
\hline
\cellcolor {blue!15} \textbf{YPS09} &  && & & &\cellcolor {blue!50}z=3.47 \\
\cellcolor {blue!15} & && & &&\cellcolor {blue!50}p=0.0003\\
\hline
\end{tabular}
}
\vspace{-2mm}
\caption{\small Pairwise comparisons of seven approaches' conversion rates, domain=``books''.}\vspace{-2mm}
\label{tab:userstudy-z-p-books}
\end{table}

\begin{table}[h]
\centering
\scriptsize
{
\setlength{\tabcolsep}{0.35em}
\begin{tabular}{|l|l|l|l|l|l|l|}
\hline
\rowcolor {blue!50}
 & \textbf{Tight} & \textbf{Diverse} & \textbf{Freebase} & \textbf{Experts} & \textbf{YPS09} & \textbf{Graph} \\ \hline
\cellcolor {blue!15}\textbf{Concise} & z=$-$0.16 &z=0.92 & \cellcolor {blue!50} z=1.49 &z=$-$0.45 &z=0.29 &z=0.14 \\
\cellcolor {blue!15} & p=0.4364 &p=0.1788 & \cellcolor {blue!50} p=0.0681 &p=0.3264 &p=0.3859 &p=0.4443 \\ \hline
\cellcolor {blue!15} \textbf{Tight} & &z=1.06 & \cellcolor {blue!50} z=1.61 &z=$-$0.28 &z=0.45 &z=0.29 \\
\cellcolor {blue!15} & & p=0.1446 & \cellcolor {blue!50}p=0.0537 &p=0.3897 &p=0.3264 &p=0.3859 \\
\hline
\cellcolor {blue!15} \textbf{Diverse} &  & & z=0.66 & \cellcolor {blue!15}z=$-$1.34 &z=$-$0.63 &z=$-$0.73 \\
\cellcolor {blue!15} & &                   &p=0.2546 &\cellcolor {blue!15}p=0.0901 &p=0.2643 &p=0.2327  \\
\hline
\cellcolor {blue!15} \textbf{Freebase} &  & & &\cellcolor {blue!15}z=$-$1.86 &z=$-$1.23 &\cellcolor {blue!15}z=$-$1.31 \\
\cellcolor {blue!15} & & & &\cellcolor {blue!15}p=0.0314 & p=0.1093 &\cellcolor {blue!15}p=0.0951 \\
\hline
\cellcolor {blue!15} \textbf{Experts} &  & &  & &z=0.73 &z=0.55 \\
\cellcolor {blue!15} & & & & & p=0.2327 & p=0.2912 \\
\hline
\cellcolor {blue!15} \textbf{YPS09} &  && & & &z=$-$0.13 \\
\cellcolor {blue!15} & & & & & & p=0.4483 \\
\hline
\end{tabular}
}
\vspace{-2mm}
\caption{\small Pairwise comparisons of seven approaches' conversion rates, domain=``film''.}\vspace{-2mm}
\label{tab:userstudy-z-p-film}
\end{table}

\begin{table}[h]
\centering
\scriptsize
{
\setlength{\tabcolsep}{0.35em}
\begin{tabular}{|l|l|l|l|l|l|l|}
\hline
\rowcolor {blue!50}
 & \textbf{Tight} & \textbf{Diverse} & \textbf{Freebase} & \textbf{Experts} & \textbf{YPS09} & \textbf{Graph} \\ \hline
\cellcolor {blue!15}\textbf{Concise} & z=$-$0.14 &\cellcolor {blue!15}z=$-$1.74 & z=0.40 &z=$-$1.01 &\cellcolor {blue!15}z=$-$2.40 &z=0.24 \\
\cellcolor {blue!15} & p=0.4443 &\cellcolor {blue!15}p=0.0409 &p=0.3446 &p=0.1562 &\cellcolor {blue!15}p=0.0082 &p=0.4052 \\
 \hline
\cellcolor {blue!15} \textbf{Tight} & &\cellcolor {blue!15}z=$-$1.57 & z=0.52 &z=$-$0.85 &\cellcolor {blue!15}z=$-$2.21 &z=0.37 \\
\cellcolor {blue!15} &  &\cellcolor {blue!15}p=0.0582 &p=0.3015  &p=0.1977 &\cellcolor {blue!15}p=0.0136 &p=0.3557 \\
\hline
\cellcolor {blue!15} \textbf{Diverse} &  & & \cellcolor {blue!50}z=2.01 &z=0.73 &z=$-$0.65 &\cellcolor {blue!50}z=1.82 \\
\cellcolor {blue!15} & & &\cellcolor {blue!50}p=0.0222 &p=0.2327 &p=0.2578 &\cellcolor {blue!50}p=0.0344 \\
\hline
\cellcolor {blue!15} \textbf{Freebase} &  & & &\cellcolor {blue!15}z=$-$1.33 &\cellcolor {blue!15}z=$-$2.61 &z=$-$0.14 \\
\cellcolor {blue!15} & & & &\cellcolor {blue!15}p=0.0918 &\cellcolor {blue!15}p=0.0045 &p=0.4443 \\
\hline
\cellcolor {blue!15} \textbf{Experts} &  & &  & &\cellcolor {blue!15}z=$-$1.38 &z=1.16 \\
\cellcolor {blue!15} & & & & &\cellcolor {blue!15}p=0.0838 &p=0.1230\\
\hline
\cellcolor {blue!15} \textbf{YPS09} &  && & & &\cellcolor {blue!50}z=2.40 \\
\cellcolor {blue!15} & && & &&\cellcolor {blue!50}p=0.0082\\
\hline
\end{tabular}
}
\vspace{-2mm}
\caption{\small Pairwise comparisons of seven approaches' conversion rates, domain=``TV''.}\vspace{-2mm}
\label{tab:userstudy-z-p-tv}
\end{table}

\begin{table}[h]
\centering
\scriptsize
{
\setlength{\tabcolsep}{0.35em}
\begin{tabular}{|l|l|l|l|l|l|l|}
\hline
\rowcolor {blue!50}
 & \textbf{Tight} & \textbf{Diverse} & \textbf{Freebase} & \textbf{Experts} & \textbf{YPS09} & \textbf{Graph} \\ \hline
\cellcolor {blue!15}\textbf{Concise} &\cellcolor {blue!15}z=$-$1.37 &z=1.16 & z=$-$1.19 &z=$-$1.15 &\cellcolor {blue!15}z=$-$1.73 &z=0.76 \\
\cellcolor {blue!15}  &\cellcolor {blue!15}p=0.0853 &p=0.1230 &p=0.1170 &p=0.1251 & \cellcolor {blue!15}p=0.0418 &p=0.2236 \\
 \hline
\cellcolor {blue!15} \textbf{Tight} & &\cellcolor {blue!50}z=2.43 & z=0.15 &z=0.22 &z=$-$0.34 &\cellcolor {blue!50}z=1.98 \\
\cellcolor {blue!15} & &\cellcolor {blue!50}p=0.0075 & p=0.4404 &p=0.4129 &p=0.3669 &\cellcolor {blue!50}p=0.0239\\
\hline
\cellcolor {blue!15} \textbf{Diverse} &  & &\cellcolor {blue!15} z=$-$2.25 &\cellcolor {blue!15}z=$-$2.23 &\cellcolor {blue!15}z=$-$2.78 &z=$-$0.34 \\
\cellcolor {blue!15} & &  &\cellcolor {blue!15}p=0.0122 &\cellcolor {blue!15}p=0.0129 &\cellcolor {blue!15}p=0.0027 &p=0.3669 \\
\hline
\cellcolor {blue!15} \textbf{Freebase} &  & & &z=0.06 &z=$-$0.48 &\cellcolor {blue!50}z=1.82 \\
\cellcolor {blue!15} & & & & p=0.4761 & p=0.3156 &\cellcolor {blue!50}p=0.0344  \\
\hline
\cellcolor {blue!15} \textbf{Experts} &  & &  & &z=$-$0.56 &\cellcolor {blue!50}z=1.79 \\
\cellcolor {blue!15} & & & & & p=0.2877 &\cellcolor {blue!50}p=0.0367\\
\hline
\cellcolor {blue!15} \textbf{YPS09} &  && & & &\cellcolor {blue!50}z=2.31 \\
\cellcolor {blue!15} & && & &&\cellcolor {blue!50}p=0.0104\\
\hline
\end{tabular}
}
\vspace{-2mm}
\caption{\small Pairwise comparisons of seven approaches' conversion rates, domain=``people''.}\vspace{-2mm}
\label{tab:userstudy-z-p-people}
\end{table}

\begin{table}[h]
\centering
\scriptsize
    \begin{tabular}{|l|l|l|l|l|}
    \hline
    \textbf{System} & \textbf{Q1} & \textbf{Q2} & \textbf{Q3} & \textbf{Q4}\\ \hline
    \textbf{Concise} & 3.5 & 4.0769 & 3.9231 & 3.6154 \\
    \textbf{Tight} & 3.5833 & 3.9167 & 4 & 3.3333 \\
    \textbf{Diverse} & 3.9231  & 3.8462 & 4.0769 & 3.6364 \\
    \textbf{Freebase} & 3.8182 & 4.0909 & 4 & 3.6  \\
    \textbf{Experts} & 3.3333 & 3.75 & 4.2727 & 3.5 \\
    \textbf{YPS09} & 3.75 & 3.8333 & 3.8462 & 3.5385 \\
    \textbf{Graph} & 4.4 & 4.1 & 4.1 & 3.3333 \\ \hline
    \end{tabular}
\vspace{-2mm}
\caption{\small Responses to user experience questions, domain=``books''.}\vspace{-2mm}
\label{tab:experience-score-book}
\end{table}

\begin{table}[h]
\centering
\scriptsize
    \begin{tabular}{|l|l|l|l|l|}
    \hline
    \textbf{System} & \textbf{Q1} & \textbf{Q2} & \textbf{Q3} & \textbf{Q4}\\ \hline
    \textbf{Concise} & 4 & 4.0909 & 4.4167 & 3.7692 \\
    \textbf{Tight} & 4.0833 & 4.6667 & 4.5 & 3.75 \\
    \textbf{Diverse} & 4.1538 & 4.4615 & 4.4615 & 3.3846 \\
    \textbf{Freebase} & 4.1818 & 4.3636 & 4.2727 & 3.4545 \\
    \textbf{Experts} & 4 & 4.0833 & 4.25 & 3.2727 \\
    \textbf{YPS09} & 3.5385 & 4.3077 & 4.2308 & 4 \\
    \textbf{Graph} & 3.8 & 4.7 & 4.6 & 4\\ \hline
    \end{tabular}
\vspace{-2mm}
\caption{\small Responses to user experience questions, domain=``film''.}\vspace{-2mm}
\label{tab:experience-score-film}
\end{table}

\begin{table}[h]
\centering
\scriptsize
    \begin{tabular}{|l|l|l|l|l|}
    \hline
    \textbf{System}   & \textbf{Q1} & \textbf{Q2} & \textbf{Q3} & \textbf{Q4}\\ \hline
    \textbf{Concise}    & 3.8462   & 3.8462 & 4.1538 & 3.5833   \\
    \textbf{Tight}   & 3.6667    & 3.8333  &  4.0833 & 3.75    \\
    \textbf{Diverse}    & 3.75    & 3.75  &3.9167 & 3     \\
    \textbf{Freebase}   & 3.8182 & 4.2727 & 4.4545 & 3.5455  \\
    \textbf{Experts}   & 4.1667   & 4.1667  & 4.5 & 4.3333   \\
    \textbf{YPS09} & 4.3077 &4.5385 & 4.4615 &  3.8333\\
    \textbf{Graph} &3.6 & 4.6 & 4.5 & 3.9\\ \hline
    \end{tabular}
\vspace{-2mm}
\caption{\small Responses to user experience questions, domain=``music''.}\vspace{-2mm}
\label{tab:experience-score-music}
\end{table}

\begin{table}[h]
\centering
\scriptsize
    \begin{tabular}{|l|l|l|l|l|}
    \hline
    \textbf{System} & \textbf{Q1} & \textbf{Q2} & \textbf{Q3} & \textbf{Q4}\\ \hline
    \textbf{Concise} & 3.7692 & 4 & 3.7692 & 3.7692 \\
    \textbf{Tight} & 4.1667 & 4.1667 & 4.1667 & 3.6667 \\
    \textbf{Diverse} & 4.0833 & 4.25 & 4.4167 & 3.6667 \\
    \textbf{Freebase} & 4.5455 & 4.3636 & 4.2727 & 3.2727 \\
    \textbf{Experts} & 4.1667 & 3.8333 & 3.8333 & 3.6667 \\
    \textbf{YPS09} & 3.5385 & 3.6154 & 3.7692 & 3 \\
    \textbf{Graph} & 3.5 & 4.6 & 4.4 & 3.9 \\ \hline
    \end{tabular}
\vspace{-2mm}
\caption{\small Responses to user experience questions, domain=``TV''.}\vspace{-2mm}
\label{tab:experience-score-tv}
\end{table}

\begin{table}[h]
\centering
\scriptsize
    \begin{tabular}{|l|l|l|l|l|}
    \hline
    \textbf{System} & \textbf{Q1} & \textbf{Q2} & \textbf{Q3} & \textbf{Q4}\\ \hline
    \textbf{Concise} & 4.2308 & 4.3846 & 4.3077 & 4 \\
    \textbf{Tight} & 2.9167 & 3.6364 & 3.4545 & 2.9167 \\
    \textbf{Diverse} & 4.0833 & 4.1667 & 4.0833 & 3.5833 \\
    \textbf{Freebase} & 3.9091 & 4.0909 & 4.0909 & 3.4545 \\
    \textbf{Experts} & 3.9167 & 4.0833 & 4.0833 & 3.75 \\
    \textbf{YPS09} & 4.3333 & 4.4615 & 4.6923 & 4.3846 \\
    \textbf{Graph} & 4.5 & 4.1 & 4 & 3.1 \\ \hline
    \end{tabular}
\vspace{-2mm}
\caption{\small Responses to user experience questions, domain=``people''.}\vspace{-2mm}
\label{tab:experience-score-people}
\end{table}

\begin{table}[h]
\centering
\scriptsize
    \begin{tabular}{|l|l|l|l|l|l|}
    \hline
    \textbf{K} & \textbf{books} & \textbf{film} & \textbf{music} & \textbf{TV} &\textbf{people} \\ \hline
    1 & 1 & 1 & 1 & 1 & 1 \\
    2 & 0.5 & 0.5  & 1 & 1 & 1 \\
    3 & 0.334 &	0.334 & 1 & 1 & 0.664 \\
    4 & 0.25 & 0.5 & 1 & 0.75 & 0.5 \\
    5 & 0.2 & 0.6 &	1 & 0.6	& 0.6 \\
    6 & 0.333 & 0.5 & 0.833 & 0.5 & 0.5 \\  \hline
    \end{tabular}
\vspace{-1mm}
\caption{\small Precision-at-$K$ of key attribute scoring in ``Freebase'', using ``Experts'' as ground truth.}\vspace{-2mm}
\label{tab:precision-at-k-gold}
\end{table}

\begin{table}[h]
\centering
\scriptsize
    \begin{tabular}{|l|l|l|l|l|l|}
    \hline
    \textbf{K}   & \textbf{books} & \textbf{film} & \textbf{music} & \textbf{TV} &\textbf{people} \\ \hline
    1 & 1 & 1 & 1 & 1 & 1 \\
    2 & 1 & 0.5 & 1 & 1 & 0.5 \\
    3 & 0.667 & 0.667 & 1 & 0.667 & 0.667 \\
    4 & 0.5 & 0.75 & 1 & 0.75 & 0.75 \\
    5 & 0.4 & 0.6 & 0.8 & 0.6 & 0.6 \\
    6 & 0.333 & 0.5 & 0.833 & 0.5 & 0.5 \\  \hline
    \end{tabular}
\vspace{-1mm}
\caption{\small Precision-at-$K$ of key attribute scoring in ``Experts'', using ``Freebase'' as ground truth.}\vspace{-2mm}
\label{tab:precision-at-k-experts}
\end{table}

\end{document}